\documentclass[ twocolumn,
aps,prd,   
               preprintnumbers,numbers,sort&compress,nofootinbib,
                            showpacs,
               colorlinks,
               linkcolor=blue,   
               citecolor=blue]{revtex4-1}
    \newcommand{\exclude}[1]{}

\usepackage{graphicx,amsmath,amssymb,bm}
\usepackage{psfrag}
\usepackage{feynmp}
\usepackage{hyperref}
\usepackage{enumitem}
\usepackage{slashed}
\usepackage{subcaption}

\def\<{\langle}
\def\>{\rangle}

\def\+{\dagger}

\def\U1A{U(1)$_{\rm A}$}

 \def\<{\langle}
\def\>{\rangle}

\def\+{\dagger}

\def\ra{\rangle}
\def\la{\langle}
\def\U1A{U(1)$_{\rm A}$}

\newcommand{\be}{\begin{eqnarray}}
\newcommand{\ee}{\end{eqnarray}}
\newcommand{\beq}{\begin{equation}}
\newcommand{\eeq}{\end{equation}}

\begin{document}

\title{The cosmological $\cal{CP}$ odd axion field as the coherent Berry's phase of the Universe}
%
%

\author{ Shuailiang Ge}
\author{ Xunyu Liang}
\author{Ariel Zhitnitsky}
\affiliation{Department of Physics and Astronomy, University of
  British Columbia, Vancouver,  Canada}
  
\begin{abstract}
We consider  a  
dark matter (DM) model     offering  a very natural  explanation of the observed relation,  $\Omega_{\rm dark} \sim  \Omega_{\rm visible}$.
  This generic consequence of  the model  is a result 
of the common   origin of  both types of matter (DM and visible) which are formed     during  the same  QCD transition.
The masses of both types of matter in this framework are proportional to one and the same   dimensional parameter of the system,    $\Lambda_{\rm QCD}$. 
The focus of the present work is the detail study  of the dynamics of the  $\cal{CP}$-odd coherent axion field $a(x)$ just before the QCD transition.  
We argue that the baryon charge separation effect  on the largest possible scales   inevitably  occurs 
as a result of merely   existence of the  coherent axion field
  in early Universe. It  leads  
to preferential formation of one species of nuggets on the  scales of the visible Universe where the axion field $a(x)$ is coherent. A natural outcome of  this preferential evolution   is that only one type of the visible baryons   remains in the system after the nuggets complete their formation.   This represents a specific mechanism on how the  baryon charge separation mechanism (when the Universe is neutral, but visible part of matter consists   the baryons only)  replaces the conventional ``baryogenesis" scenarios.  
 \end{abstract}
\vspace{0.1in}

\maketitle
\section{Introduction}
\label{intro}
The nature of dark matter (DM) and the asymmetry between matter and antimatter    are generally assumed to be two unrelated open questions in cosmology. However, we advocate a model, originally suggested in 
 \cite{Zhitnitsky:2002qa,Oaknin:2003uv}, that these  two  fundamental, naively unrelated,  questions are,  in fact,  closely interconnected. In this model, the matter-antimatter asymmetry (the so-called baryogenesis) is just a $\cal CP$ violating charge separation process which occurs as a result of a coherent    $\cal CP$-odd axion field over the whole Universe.  The unobserved antibaryons in this framework come to comprise the DM in form of the dense heavy nuggets, similar to the Witten's strangelets \cite{Witten:1984rs}.

This work is the continuation of our previous studies  \cite{Liang:2016tqc}  with  the   main focus on   evolution of    a single  nugget.    A related but distinct question  on the global $\cal CP$ violating separation of baryonic charge was mentioned in  \cite{Liang:2016tqc} without any quantitative computations. 

\begin{figure}
\centering
\captionsetup{justification=raggedright}
\includegraphics[width=0.8\linewidth]{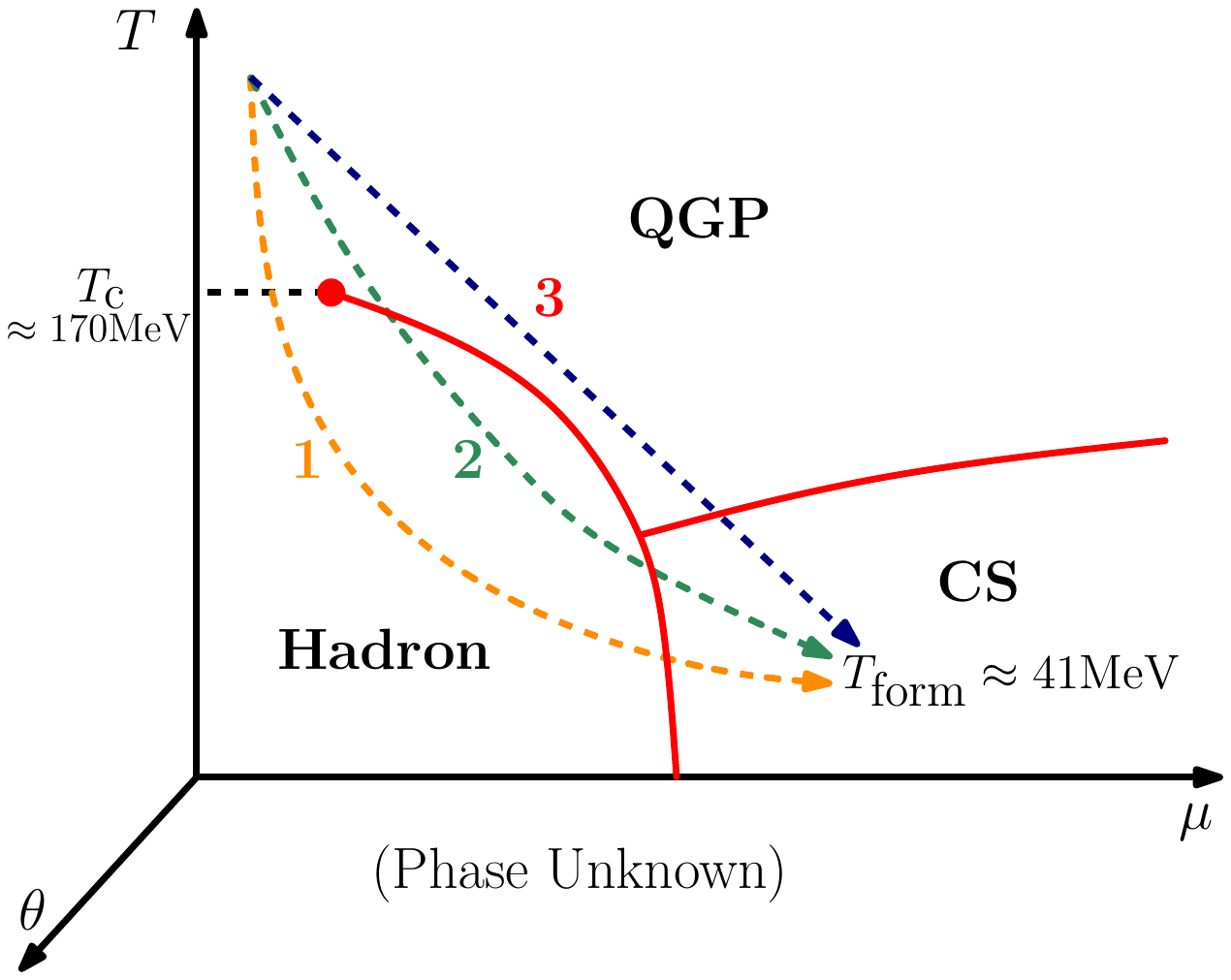}
\caption{The conjectured phase diagram.  The plot is taken from \cite{Liang:2016tqc}. Possible cooling paths are denoted as path 1, 2 or  3. The phase diagram is in fact much more complicated as the dependence on the third essential parameter, the $\theta$ is not shown as it is largely unknown. It is assumed that the final destination of the nuggets   is the CS region with $T_{\rm form}\approx 41$ MeV, $\mu >\mu_c$ and $\theta\approx 0$.   }
\label{fig:phase_diagram}
\end{figure}

The main goal  of  the present work is to present  robust arguments (supported by the detail analytical and numerical  computations)  that a sufficiently strong global $\cal{CP}$ violation   in form  of the fundamental axion field $\theta(x)$ inevitably leads to such separation of baryon charges. This phenomena of separation  is  preceding  the QCD transition\footnote{\label{QCD_phase}It is known that 
the QCD transition is actually a crossover rather than a phase transition \cite{Aoki:2006we} at $\theta=0$. At a non vanishing $\theta\neq 0$ 
the phase diagram is not known. However,  
in context of the present paper the important factor is the scale $T_c\sim 170$ MeV where transition  happens rather than its precise nature.
This region on the QCD phase diagram  is denoted by blue dashed  line  in vicinity of $T_c$ shown  on Fig.\ref{fig:phase_diagram}}.  As we argue in the present work this axion $\theta(x)$ field  can be thought  as the Berry's phase which is coherently accumulated  on the largest possible scales of the visible Universe. Precisely this coherence leads to a preferential evolution  of the nuggets when  one type of the visible baryons (not anti- baryons) prevails in the system.   This source of strong $\cal CP$ violation is no longer available at the present epoch as a result of the axion dynamics, see original papers \cite{axion,KSVZ,DFSZ}, recent reviews   
  \cite{vanBibber:2006rb, Asztalos:2006kz,Sikivie:2008,Raffelt:2006cw,Sikivie:2009fv,Rosenberg:2015kxa,Graham:2015ouw,Ringwald2016} and recent results/proposals on the axion search experiments \cite{Rybka:2014cya,Budker:2013hfa,Graham:2013gfa,Sikivie:2013laa,Beck,Stadnik:2013raa,Sikivie:2014lha,McAllister:2015zcz,Hill:2015kva,Hill:2015vma,Kahn:2016aff,Barbieri:2016vwg,Arvanitaki:2014dfa}. 
   
 \exclude{  
While a basic review of the observational consequences and the crucial ingredients will be presented in Sec. \ref{nuggets} and \ref{formation} respectively,  a few essential remarks will  be made  below.
}
The basic consequence of this framework is that the visible and dark matter densities are  of the same order of magnitude  \cite{Liang:2016tqc}:
\be
\label{Omega}
 \Omega_{\rm dark} \approx \Omega_{\rm visible}.
\ee
This is a very generic consequence of the entire framework, and it  is not sensitive\footnote{\label{m_a}The axion's mass $m_a (T)$ as a function of the temperature at $T>T_c$ 
has been computed using  the lattice simulations by different groups \cite{Kitano:2015fla,Bonati:2015vqz,Borsanyi:2016ksw,Petreczky:2016vrs} with somewhat contradicting results. As we argue in  Sec. \ref{difference} our main claim  (\ref{Omega})  is  insensitive to a precise value of  the axion mass $m_a$ }   to the parameters of the system, such as the axion mass $m_a$.   In our framework the  relation (\ref{Omega})  emerges in a very natural way  because 
both types of matter (visible and dark) are proportional to a single dimensional parameter of the system, $\Lambda_{\rm QCD}$. 

The very generic relation (\ref{Omega}) of this framework is also not sensitive to the 
  initial value  $\theta_0$ of the  
  coherent axion field when it starts to oscillate.     This should be 
 contrasted with conventional mechanisms (such as production of the axions due to the  misalignment mechanism or due to the  domain wall network decay) which are highly sensitive to these parameters as $\Omega_{\rm axion}  \sim \theta^2_0   m_a^{-7/6}$, see reviews  \cite{vanBibber:2006rb, Asztalos:2006kz,Sikivie:2008,Raffelt:2006cw,Sikivie:2009fv,Rosenberg:2015kxa,Graham:2015ouw,Ringwald2016}.

 In particular, the observed  ratio $\Omega_{\rm dark}\simeq5\cdot\Omega_{\rm visible}$ would  correspond to a slight asymmetric excess of anti-nuggets comparing to the nuggets by a factor of $\sim0.5$ at the end of the nugget's formation, if one assumes that the nuggets saturate the dark matter density today. Thus, the approximate  observed ratio  $\Omega_{\rm dark}\simeq5\cdot\Omega_{\rm visible}$ roughly corresponds to
\be
\label{ratio1}
|B_{\rm visible}|: |B_{\rm nuggets}|: |B_{\rm antinuggets}|\simeq 1:2:3, 
\ee
such that the net baryonic charge is zero. 

\exclude{To reiterate, if these two processes are not fundamentally related to each other, the two components $\Omega_{\rm dark}$ and $\Omega_{\rm visible}$ could easily exist at vastly different scales. }

The quark nuggets at zero temperature satisfy the main criteria to be the DM candidates as they are absolutely stable  configurations made of quarks and gluons surrounded by the axion domain wall 
as described in the original paper \cite{Zhitnitsky:2002qa}. However, 
unlike conventional dark matter candidates, such as WIMPs, the dark-matter/antimatter
nuggets   are macroscopically large objects and they strongly interact with visible matter.   The 
quark nuggets  do not contradict the   observational
constraints on dark matter or
antimatter  for three main reasons~\cite{Zhitnitsky:2006vt}:
\begin{itemize} 
\item They carry  very large baryon charge 
$|B|  \gtrsim 10^{25}$, and so their number density is very small $\sim B^{-1}$.  
 As a result of this unique feature, their interaction  with visible matter is highly  inefficient, and 
therefore, the nuggets are perfectly qualify  as  DM  candidates. In particular, the quark nuggets  essentially decouple 
from CMB photons, and therefore, they do not destroy conventional picture for the structure formation; 
\item The core of the  nuggets have nuclear densities. Therefore, the relevant  effective interaction
is very small $\sigma/M \sim 10^{-10}$ ~cm$^2$/g. Numerically, it is  comparable with conventional WIMPs values.
Therefore, it is consistent  with the typical astrophysical
and cosmological constraints which  are normally represented as 
$\sigma/M<1$~cm$^2$/g;
\item The quark nuggets have  very  large binding energy due to the   large    gap $\Delta \sim 100$ MeV in superconducting phases.  
Therefore, the baryon charge is so strongly bounded in the core of the nugget that  it  is not available to participate in big bang nucleosynthesis
(\textsc{bbn})  at $T \approx 1$~MeV, long after the nuggets had been formed. 
\end{itemize} 
We emphasize that the weakness of the visible-dark matter interaction 
in this model is due to a  small geometrical parameter $\sigma/M \sim B^{-1/3}$ 
  which replaces 
the conventional requirement of sufficiently weak interactions for WIMPs. 

   We conclude this Introduction  with the following remark. We consider the model which has a single fundamental parameter (the mean baryon number of a nugget $\la B\ra\sim10^{25}$, corresponding to the axion mass $m_a\simeq10^{-4}~\rm eV$). It has been shown 
    that this model is consistent with  all known 
   observations,  including the satellite and ground based constraints.  It has been also shown that there is a number of frequency bands where some excess of emission was observed, and this model may explain some portion, or even entire excess of the observed radiation in these frequency bands.  We refer to recent short review \cite{Zhitnitsky:2016cir}  with large number of references  on   original   computations which have been carried out for  each specific frequency band where some excess of radiation has been observed.

   The paper is organized as follows.   
  In  Section  \ref{formation}  we overview the  big picture of our framework when the ``baryogenesis" is replaced by the baryon charge separation
  scenario, and the DM is represented by quark nuggets and anti-nuggets. We list   the  crucial  ingredients of the entire framework  by paying   special attention to the role  the coherent $\cal{CP}$-odd axion field discussed in details in subsection \ref{CP}.    
  
  Essentially, the main objective of the present  work is to elaborate  on this specific key  element of the proposal which has not received sufficient attention in the previous paper \cite{Liang:2016tqc}. Our goal of this work is  to present solid quantitative computations suggesting  that this  coherent $\cal{CP}$-odd axion field generates the  disparity  between nuggets and anti-nuggets. This asymmetry  automatically leads to the    relation  (\ref{Omega}) which we claim is very generic consequence of the entire framework. 
   
   The readers who  are not interested in any technical details may skip the next two sections  (\ref{sec:evolution}  and \ref{difference})
   and jump directly to the concluding section \ref{Conclusion}.
   
     In Section \ref{sec:evolution} we argue that the nuggets and anti-nuggets behave in a drastically different way as a result of interaction with this $\cal{CP}$-odd coherent   cosmological axion field. 
   Finally, in section \ref{difference} we argue that the difference in evolution of the nuggets and anti-nuggets is always of  order of one effect, being  insensitive to initial conditions nor to  the dynamical  parameters of the system. As a result, the main claim   of this proposal represented by eq.(\ref{Omega}) is very robust consequence of the framework and is not a  result of any fine tuning adjustments.

\section{Big picture and  the  crucial  elements of the  proposal }\label{formation}

In this section we summarize the  crucial elements of the proposal which describe the formation   of the nuggets. These ingredients determine the basic properties of the nuggets, such as the size, local accretion of baryonic charge, abundance, stability, and global $\cal CP$ violating charge separation leading to the disparity between nuggets and anti-nuggets.  Most of these   basic elements of the proposal have  been    discussed previously in the original papers \cite{Zhitnitsky:2002qa} and  \cite{Liang:2016tqc}.  We include them into the present work   to make it   self-contained.   One crucial ingredient of the proposal which was mentioned in  \cite{Liang:2016tqc} but has not been fully elaborated there   is related to the $\cal CP$ violating processes leading to the   asymmetry between nuggets and anti-nuggets. We highlight the basic idea on $\cal CP$ violation  in subsection \ref{CP}, while the detail  analysis 
of this ingredient  of the proposal is carried out in  ``technical" sections  \ref{sec:evolution}  and \ref{difference}.

\subsection{$N_{\rm DW}=1$ domain walls}\label{DW}

The first important element, axion domain wall \cite{Sikivie,Vilenkin}, determines the size of a nugget as originally suggested in \cite{Zhitnitsky:2002qa}. The axion field $\theta$ is an angular variable, and therefore supports various types of the topological domain walls (DW), including the so-called $N_{\rm DW}=1$ domain walls when $\theta$ interpolates between one and the same physical vacuum state with the same energy $\theta\rightarrow\theta+2\pi n$.     

It is important to emphasize that while the axion string formation happens during the Peccei-Quinn (PQ)  phase transition, the domain wall formation occurs at  the QCD temperature at $T\sim 1$ GeV when  the axion potential gets tilted.  In other words, the formation of the DW-string network is the two stages process, rather than a single event, see  \cite{Vilenkin} for review.  Furthermore, the DW energy density per unit volume   is characterized by a  typical QCD scale, rather than the PQ scale. Therefore, the closed DW surfaces, without attached strings,  could be formed  during the QCD transition, though the number density of such objects  is    suppressed  with increasing the  size  of the objects \cite{Vilenkin,Chang:1998tb}, see subsection \ref{KZ}  with few more comments and estimates  on this suppression.  

One should also add that the numerical simulations \cite{Chang:1998tb} support this picture by observing the formation of the 
large DW at the QCD temperature and their decay due to the attached strings. The closed DW  without attached strings have been  also observed in  numerical simulations  \cite{Vilenkin,Chang:1998tb}, though the probability to find the closed walls  is suppressed according to  (\ref{KZ1}).
 Due to this suppression, the role of these closed DW is normally  ignored in the analysis of the DW decays to the DM axions. However, precisely these closed DW surfaces 
formed at the QCD scale play the key role in our proposal \cite{Zhitnitsky:2002qa}, see additional comments   at the end of this subsection.

One should remark here that it is normally assumed that for the topological defects to be formed,  the   PQ  phase transition must occur after inflation. This argument is valid for a generic type of  domain walls with  $ N_{DW}\neq 1$.  The conventional argument is    based on  the fact that  few physically  \textit{different vacua} with the same energy must be present inside of the same horizon for the domain walls to be formed. The $ N_{DW}=1$ domain walls
are unique and very special in the  sense that $\theta$ interpolates between  \textit{one} and the \textit{same}   physical vacuum state.   Such $ N_{DW}=1$ domain walls can be formed even if the PQ phase transition occurred before inflation and a unique physical vacuum occupies entire Universe \cite{Liang:2016tqc}. 
  
  In other words, while in our proposal the inflation is assumed to occur after the  PQ phase transition and the  axion field $\theta (t)$ is coherent  in the entire visible Universe,  nevertheless  the  $ N_{DW}=1$ closed domain walls  can  still be formed. The nonzero $\theta$ would essentially lead to a global $\cal CP$ violating separation of baryonic charge as discussed in item  \ref{CP}.

The axion domain walls normally  start to form once the axion field get tilted at temperature $T_a$. As the tilt becomes   more pronounced (at the transition when the chiral condensate forms at $T_c$) the DW formation becomes much more efficient.  We should  expect, in general, that  the $N_{\rm DW}=1$ domain walls form at any moment  between $T_a$ and $T_c$. The width  of the domain wall  depends on the mass of the axion, which would ultimately determine the size of the nugget being formed. 

One next comment is as follows.  It has been realized many years after the original publication \cite{Sikivie} that the axion domain walls generically  demonstrate a sandwich-like substructure on the QCD scale $\Lambda_{\rm QCD}^{-1}\simeq\rm fm$. Such a substructure is supported by analysis \cite{FZ} of QCD in the large $N$ limit with inclusion of the $\eta'$ field. It is also supported by analysis \cite{SG} of supersymmetric models where a similar $\theta$ vacuum  structure occurs. The same structure also occurs in colour superconducting (CS) phase where the corresponding domain walls have been explicitly constructed \cite{Son:2000fh}.

Significance of  the QCD substructure is that it is capable to  squeeze the quarks to bring the system into the  CS state, as originally suggested in \cite{Zhitnitsky:2002qa}. The fact that the CS phase  representing  the lowest energy state might be realized in nature in the core of neutron stars has been known for quite sometime.  A less known application of the CS phase   is that the axion DW with the QCD substructure may   replace the gravity and play the  role of a squeezer 
to produce an  absolutely stable quark nugget at $T=0$  as suggested in \cite{Zhitnitsky:2002qa}. 

The time evolution of these nuggets   at $T\neq 0$ is much more  involved  problem than the study of the equilibrium configurations at $T=0$ carried out in 
 \cite{Zhitnitsky:2002qa}.  The corresponding problem of the time evolution at $T\neq 0$  has been recently addressed in
\cite{Liang:2016tqc}. In particular, in that paper   it has been  shown that  the nuggets assume the equilibrium at low temperature with   the  lowest possible 
state   if the initial size of the nugget is sufficiently large.  These results fully support the earlier studies  of ref.\cite{Zhitnitsky:2002qa} 
devoted to the equilibrium configurations at $T=0$.

To conclude this subsection. The formation of the axion DW at the QCD transition is extremely generic phenomenon which  inevitably occurs as a result of 
$\theta\rightarrow\theta+2\pi n$ periodicity. These axion DW typically have the QCD substructure as mentioned above.  Furthermore, the formation of the closed surfaces at the QCD transition  (which eventually may produce the quark nuggets)    also represents a very generic feature of the system, which will be further elaborated in subsection \ref{KZ}.

\subsection{ Spontaneous  breaking of $\cal{C}$ symmetry on small scale of order $\xi(T)$ }\label{spontaneous}

  The second important    element  of the formation mechanism can be explained as follows.   There is  another substructure with a similar QCD scale  (in addition to the known   substructures  expressed in terms of the  $\eta'$ and gluon   fields as explained in section \ref{DW} above) which  carries  the baryon charge. 
   This additional substructure  is a novel feature of the axion domain walls which has been    explored  only recently in \cite{Liang:2016tqc}. As (anti)quarks being trapped in the core of  the domain wall,  the nuggets itself    slowly {\it accrete}  the  baryonic charge  as a result of evolution. Exactly this new effect is  eventually responsible for local accretion of the baryonic charge by the nugget. 
   
    Indeed, in the background of the domain wall, the physics essentially  depends on two variables, $(t, z)$. One can show  that in this circumstances the  total  baryon charge  $B$ accumulated on  a nugget   is determined  by  the degeneracy factor in vicinity of the domain wall    \cite{Liang:2016tqc}
   \be
\label{N}
B=N \cdot g\cdot \int \frac{ d^2 x_{\perp}d^2k_{\perp}}{(2\pi)^2} \frac{1}{\exp(\frac{\epsilon-\mu}{T})+1}.
\ee
In this formula the induced baryon number $N$ per degree of freedom may randomly assumed any integer  value, positive or negative; 
 the coefficient  $g$  describes the  degeneracy factor, e.g. $g\simeq N_cN_f$ in CS phase and $\mu$ is the chemical potential in the vicinity of the domain wall. Thus, the  size distribution for quark and antiquark nuggets  will be  identically the same size   if the external environment is $\cal CP$ even, which is the case when fundamental $\theta=0$.

The crucial  element here   is that the domain walls  will acquire the baryon or anti-baryon charge as a generic feature of the system. This is because the domain wall tension is mainly determined by the axion field while the QCD substructure leads to small correction factor of order $\sim\Lambda_{\rm QCD}/f_a\ll1$. Therefore, the presence of the QCD substructure with nonvanishing $N\neq0$ increases the domain wall tension only slightly. Consequently, this implies that the domain wall closed bubbles carrying the baryon or antibaryon charge will be copiously produced during the transition as they are very \textit{generic configurations of the system}. Furthermore, the baryonic charge cannot leave the system during the time evolution as it is strongly bound to the wall due to the topological reasons.  The corresponding binding energy per quark is order of $\mu$ and increases with time as shown in \cite{Liang:2016tqc}. One can interpret this phenomenon as a \textit{local spontaneous  breaking of $\cal{C}$ symmetry}, when on the scales of order the correlation length $\xi(T)\sim m_a^{-1}(T)$, the nuggets may acquire the positive or negative chemical potential $\mu$  with equal probability. This is because the sign of $N$ in Eq. (\ref{N}) may assume any positive or negative values with equal probabilities.

\subsection{Kibble-Zurek mechanism}\label{KZ} 

The Kibble-Zurek (KZ) mechanism  gives a generic picture of formation of the topological defects during a phase transition.
We refer to the original papers  \cite{KZ},  review \cite{KZ-review} and the textbook \cite{Vilenkin} for a general overview. For our specific purposes of the DW formation, the KZ mechanism suggests that once the axion potential is sufficiently tilted the $N_{\rm DW}=1$ closed domain walls form at the QCD scale.   Some time after $T_a$ the system is dominated by a single, percolated, highly folded and crumple domain wall network of very complicated topology. In addition, there will be a finite portion of the closed walls (bubbles) with typical size of order correlation length $\xi(T)$, which is defined as an average distance between folded domain walls at temperature $T$. Parametrically, the correlation length $\xi(T)\sim m^{-1}_a(T)$ is determined by the axion mass, and obviously varies with time during the network evolution.     Precisely these bubbles are capable to form the nuggets and play the crucial role in our analysis.
 It is known that the probability $n$ of finding closed walls with very large size $R\gg\xi$ is exponentially small, see  \cite{Vilenkin} for review, 
\be
\label{KZ1}
n\sim \exp{\left[ -\frac{R^2}{\xi^2(T)}\right]}.
\ee
The key point for our proposal is mere  existence of these finite closed bubbles made of the axion domain walls. Normally it is assumed that these closed bubbles collapse as a result of the domain wall pressure, and do not play any significant role in dynamics of the system because the total area of these bubbles is sufficiently small in comparison with the area of the dominant percolated domain wall network. However, as we already mentioned in section \ref{DW} some of  these closed bubbles do not collapse  due to the Fermi pressure acting inside of the bubbles. 
The equilibrium is achieved when the Fermi pressure from inside due to the degenerate quarks   equals to the pressure from outside due to the axion domain wall.
This is precisely the condition of the  stability analyzed  in ref. \cite{Zhitnitsky:2002qa}. 

There are many papers devoted to analysis of the network made of the domain walls bounded by the strings. There are also many papers devoted to the problem (and its possible resolutions) on  the domain wall dominance  of the Universe. 
We have nothing new to add to these subjects  and we refer  
 to the original literature  \cite{Chang:1998tb,Wantz:2009it,Hiramatsu:2012gg,Kawasaki:2014sqa,Fleury:2015aca} and reviews \cite{Vilenkin, Sikivie:2008} on this matter.  Those axions (along with the axions produced by  the conventional misalignment mechanism \cite{Wantz:2009it,misalignment})  will contribute to the dark matter density today. The corresponding contribution to DM density is highly sensitive to the axion mass as $\Omega_{\rm dark}\sim m_a^{-7/6}$, and it is not part of our  framework. 
 
 Instead, our proposal focuses on  the dynamics of the closed bubbles  (\ref{KZ1}), which are formed during the QCD transition. These closed bubbles are  normally ignored in computations of the axion production. Precisely these closed bubbles   will eventually become absolutely  stable quark nuggets and may serve as the dark matter candidates according to the proposal \cite{Zhitnitsky:2002qa,Liang:2016tqc}. 
 
 The efficiency of the production of these bubbles   has been estimated in ref.\cite{Liang:2016tqc}, based on 
 assumption that this mechanism saturates  the observed  ratio $n_B/n_{\gamma}\sim 10^{-10}$.
We shall not discuss  this  model-dependent estimate   in the present work  as the main goal of the present analysis  is   to demonstrate   that $c(T)\sim 1 $ in eq. (\ref{ratio2}).

\subsection{Colour Superconductivity}\label{CS}

The existence of CS phase in QCD is the crucial element for the stability of quark nuggets. In astrophysics, a CS is known to be the plausible phase in the neutron star interiors and in the violent events associated with collapse of massive stars or collision of neutron star, see review papers \cite{Alford:2007xm,Rajagopal:2000wf} on the subject.   The CS phase becomes energetically favourable when quarks are squeezed to few times of nuclear density.

Analogous to the gravitational squeezing in neutron star, the CS phase might be  realized in  quark nuggets due to the surface  tension of the axion domain wall as advocated in  \cite{Zhitnitsky:2002qa}. The domain wall bubbles after formation   will undergo a large number of    bounces  with typical frequency 
$\omega\sim m_a$ until they  settle down at the equilibrium configuration   \cite{Liang:2016tqc}.  As the temperature cools down, the  oscillations and squeezing  will turn the bulk of quarks into an equilibrium position with ground state being in the CS phase.

\exclude{Regarding to the surface tension and Fermi pressure (from quarks trapped inside), the domain wall bubbles   will undergo cycles of violent collapse and rebounce until finally settle down due to QCD viscosity. As the temperature cools down to the QCD transition, the violent oscillation will turn the bulk of quarks into an equilibrium with ground state being in a CS phase. 
As we advocated in \cite{Liang:2016tqc}, this is a very plausible fate of a relatively large bubble of size $R\sim\xi(T)$ made of the axion domain walls which were produced after the QCD transition. }
The corresponding  time evolution of an oscillating bubble can be  approximated\footnote{One should 
emphasize that simple  analytical expression (\ref{eq:6.R2}) is presented here for illustrative  purposes to demonstrate the oscillating and damping features of the nugget's evolution. Numerically,  it is only justified  at the very end of the evolution when the amplitude of the oscillations is small, and a complicated effective potential can be expanded around its minimum as discussed in  \cite{Liang:2016tqc}, see also related discussions after eq.(\ref{eq:6.R1tw}). In other words, formula (\ref{eq:6.R2}) properly describes the dynamics of the nugget only when  the  nugget's  formation is almost complete. In our numerical studies given in Appendix \ref{Appendix:CPodd_numerical} we use the original potential without assuming that the oscillations are small.} as follows:
\begin{equation}
\label{eq:6.R2}
R(t)=R_{\rm form}+(R_0-R_{\rm form})e^{-t/\tau}\cos\omega t , 
\end{equation}
where  the initial radius of a bubble $R_0\sim\xi(T)$ is assumed to be of order the correlation length $\xi(T)$. The final size $R_{\rm form}$ of a  bubble
represents the equilibrium configuration  when formation is almost complete. In formula (\ref{eq:6.R2}) parameter $\tau$ represents a typical damping time scale which is expressed in terms of the axion mass $m_a$,  and the QCD parameters such as viscosity. It turns out the numerical value of $\tau$ is of order of the cosmological scale $\tau\sim10^{-4}\rm s$. This numerical value is fully consistent with our anticipation that the temperature of the Universe drops approximately by a factor of $\sim3$ or so during the formation period. During the same period of time the chemical potential $\mu$ inside the nugget reaches sufficiently large value  when the CS phase sets in \cite{Liang:2016tqc}.

\subsection{Coherent $\cal{CP}$-odd axion field}\label{CP}

The key element to be investigated in the present work, is the $\cal CP$ asymmetric charge separation originating from the globally coherent axion field $\theta(t)$. In this subsection  we outline the basic ideas while all technical details will be elaborated in sections \ref{sec:evolution}  and \ref{difference}. 
 
It is well known that the    axion  dynamics  at sufficiently large temperature $T>T_c$  is determined by the  coherent state of axions at rest, see e.g. review paper \cite{Sikivie:2008}:
\be
\label{axion}
\theta (t)\sim \frac{C}{t^{3/4}}\cos\int^t dt' \omega_a(t'), ~ \omega_a^2(t)=m_a^2(t)+\frac{3}{16t^2}, ~~~~
\ee
where $C$ is a constant, and $t=\frac{1}{2H}$ is the cosmic time. This formula describes the dynamics of the axion  field after the moment 
$t_1$ determined by the condition $m_a(t_1) t_1=1$ when the axion mass $m_a$  effectively turns on at $t_1$.  
Precisely this moment is relevant for our studies as the domain walls start to form after $t_1$.
For sufficiently large $t$ when  $m_a(t)t\gg 1$ the second term in expression for $\omega_a^2(t) $ can be ignored and the frequency of the oscillations $\omega_a (t)$ is determined by the axion mass   at time $t$,  
\be
\label{axion1}
 \dot{\theta} (t)\sim \omega_a (t)\simeq m_a(t). 
\ee
The  key  point in what follows is the observation  that $ \theta(t)$   is one and the same  in the entire visible Universe.
 \exclude{t should be contrasted with a  system (relevant for the axion dark matter   searches \cite{vanBibber:2006rb, Sikivie:2008, Asztalos:2006kz,Raffelt:2006cw,Sikivie:2009fv,Rybka:2014cya,Rosenberg:2015kxa,Graham:2015ouw,Ringwald2016,Budker:2013hfa,Graham:2013gfa,Sikivie:2013laa,Beck,Stadnik:2013raa,Sikivie:2014lha,McAllister:2015zcz,Hill:2015kva,Hill:2015vma,Kahn:2016aff,Barbieri:2016vwg}) when  the axion field is represented by     conventional propagating  axions in which case the correlation scale is determined   by the De Broglie wave length  $ \lambda_D= {\hbar}/{(m_a v_a)}$ of a single individual axion moving with velocity $\vec{v}_a$.
}
   One should also emphasize that this assumption on coherency  of the axion field on very large scales is consistent with formation of $N_{\rm DW}=1$ domain walls as explained in subsection  \ref{DW}.

Similar to the case of formation temperature $T_{\rm form}$ discussed in details in \cite{Liang:2016tqc}, precise dynamical computation of this $\cal CP$ asymmetry due to the coherent axion field $\theta(t)$ is a hard problem of strongly coupled QCD at $\theta\neq0$ when even the phase diagram,   schematically shown on Fig.\ref{fig:phase_diagram}  is not yet known. It depends on a number of specific properties of the nuggets, their evolution, their environment, modification of the hadron spectrum at $\theta\neq0$  and corresponding cross sections as mentioned in \cite{Liang:2016tqc}. All these factors equally contribute to the difference between the nuggets and anti-nuggets. One can  effectively account for these coherent $\cal CP$  odd effects by  introducing  an unknown coefficient $c(T)$ of order one as follows  \cite{Liang:2016tqc}  
\be
\label{ratio2}
B_{\rm antinuggets}=c(T) \cdot   B_{\rm nuggets}  ,~~{\rm where  } ~~ |c(T)| \sim 1.~~
\ee

The main goal of the present work is to provide the quantitative numerical analysis supporting the basic assumption (\ref{ratio2}). We shall argue that $|c(T)|\sim1$ is indeed very likely to be of order of one as a result of the $\cal CP$ violating processes which took place coherently on enormous scale of the entire visible Universe before  the QCD transition. We shall argue below that this very generic outcome of this framework is not very sensitive  to initial conditions (such as the magnitude of $\theta_0$ at the moment of formation), nor to a precise value of the axion mass $m_a(T)$ at $T>T_c$ when the domain wall network only started to form. The  fundamental ration (\ref{Omega}) is the direct consequence of  $|c(T)|\sim1$. Therefore the arguments supporting (\ref{ratio2}) are essentially equivalent to the  basic claim of this framework (\ref{Omega}). 

 What is the significance of  eq. (\ref{ratio2})? The most important and unambiguous consequence of the
  eq. (\ref{ratio2})    is  that
 the baryon charge in form of the visible matter can be also expressed in terms of the same coefficient $c(T)\sim 1$
 as follows  
$B_{\rm visible} =- (B_{\rm antinuggets}    + B_{\rm nuggets})$.  
Using eq. (\ref{ratio2}) the expression for the visible matter  $B_{\rm visible}$ can be rewritten as \cite{Liang:2016tqc}
  \be
 \label{ratio4}
    &&B_{\rm visible}\equiv \left(B_{\rm baryons}+B_{\rm antibaryons}\right) \\
&&=  -\left[1+c(T)\right] B_{\rm nuggets} =-\left[1+\frac{1}{c(T)}\right] B_{\rm antinuggets}.   \nonumber
  \ee
  This is very important relation which we would like to represent 
    in terms of the measured observables  $\Omega_{\rm visible}$ and $\Omega_{\rm dark}$ at late times  when 
    the visible matter consists   the baryons only  \cite{Liang:2016tqc}:
  \be
  \label{ratio_omega}
  \Omega_{\rm dark}\simeq \left(\frac{1+|c(T)|}{\left|1+c(T)\right|}\right)\cdot \Omega_{\rm visible} ~~ {\rm at} ~~ T\leq T_{\rm form}.
  \ee
  Important comment here is that 
  the relation (\ref{ratio4}) holds as long as the thermal equilibrium is maintained.
Furthermore, the thermal equilibrium implies   that each individual contribution  $|B_{\rm baryons}|\sim |B_{\rm antibaryons}|$ entering  (\ref{ratio4})     is  many orders of magnitude greater  than the baryon charge hidden in the form of the nuggets and anti-nuggets at  earlier  times when $T_c>T> T_{\rm form}$. However,    the net baryon charge  which is labeled as $B_{\rm visible}$   in eq. (\ref{ratio4}) is the same order of magnitude  as the net baryon charge hidden in the form of the nuggets and anti-nuggets.
 
For a specific value of   $c(T_{\rm form})\simeq -1.5$ the relations (\ref{ratio4}) and (\ref{ratio_omega}) assume the form
\be
\label{ratio5}
B_{\rm visible}&\simeq& \frac{1}{2} B_{\rm nuggets}\simeq -\frac{1}{3}B_{\rm antinuggets},  \nonumber\\ \Omega_{\rm dark}&\simeq &5\cdot\Omega_{\rm visible}. 
\ee
These numerical values    coincide with approximate relation  (\ref{ratio1}) presented in Introduction. 
 The coefficient $\sim 5$ in relation  $\Omega_{\rm dark}\simeq 5\cdot\Omega_{\rm visible}$ is obviously not universal, but relation 
 (\ref{ratio_omega}) is universal, and 
  very generic consequence of the entire framework, which   was the main motivation for the proposal \cite{Zhitnitsky:2002qa,Oaknin:2003uv}.

  We shall argue in next sections \ref{sec:evolution} and \ref{difference} that $|c(T)|\sim 1$ is indeed of  order of one. Furthermore, we shall argue this feature of the system  is universal, and not very sensitive to the axion mass $m_a(T)$ nor to the initial value of  $\theta_0$  when  the  bubbles  start to oscillate and slowly accrete the baryon charge. The only crucial factor in our arguments is that    the axion field  $\theta(t)$ can be represented by the coherent superposition of the axions at rest (\ref{axion}).

\section{Evolution of the (anti) nuggets   in the background of the   axion field}
\label{sec:evolution}
In this section, we study the profound consequences of the coherent axion phase on the nugget's formation. 
 Essentially,  the main goal of this   section is to present analytical and numerical arguments    suggesting  that small  $\mathcal{CP}$ violating effects will produce a   large disparity  (of order of one) between  the  nuggets and  anti-nuggets.  

We start with subsection   \ref{symmetry} where we  
present some qualitative explanation of how a relatively small fundamental  coupling between quarks/antiquarks   and the   coherent axion $\theta(t)$ field given by (\ref{axion}) may, nevertheless,  produce a  large asymmetry  in properties between the nuggets and anti-nuggets.  
  In following  subsections \ref{analytical} and  \ref{qualitative}  we develop the technical tools to address these questions, while in   subsection \ref{slow}   we   present our estimates supporting the  main claim of this work  that $|c(T)|\sim 1$.  
     
\subsection{\label{symmetry}Qualitative analysis  }
 Before we start our specific quantitative studies     in this section we want to formulate few generic relations 
 which characterize the system and which depend exclusively   on symmetric features  of the system, rather than on some   specific dynamical model-dependent computations which follow.  
 
 First of all, let us remind  that if $\la\theta\ra=0$  the baryon charge hidden in nuggets on average is equal to the baryon charge hidden in  anti-nuggets, of course with sign minus. 
Indeed, the studies  of the anti-nuggets can be achieved by  flipping  the sign of  the chemical potential $\mu\rightarrow -\mu$ in analysis of ref. \cite{Liang:2016tqc}.  One can restore the original form of the effective  Lagrangian  by  flipping    the sign for the axion $\theta$ field as discussed in \cite{Liang:2016tqc}.  
  These  symmetry arguments imply that as long as the  pseudo-scalar  axion field $\theta$ fluctuates around zero $\la\theta\ra=0$ as conventional pseudo-scalar fields (such as $\pi, \eta'$ mesons, for example), the theory remains invariant under $\cal {P}$ and $\cal {CP}$ symmetries and on average an equal number of nuggets and anti-nuggets    carrying  equal baryon (anti-baryon) charges would form.  
  
  However, if $\la\theta\ra\neq 0$ there are many  strong   processes   taking place inside and outside the nuggets (such as annihilation, evaporation, scattering, etc) which are slightly different\footnote{\label{processes}In particular, a  slight difference of the ground states characterized by the condensate (\ref{result}) leads to the different properties of the quasiparticles and scattering amplitudes inside and outside the nuggets due to $\mu$ dependence of the $\theta$ dependent condensate (\ref{result}). The same effects also contribute to  some disparity  in  transmission/reflection coefficients, minor  differences in  viscosity, annihilation  and evaporation rates, etc} for nuggets and anti-nuggets, as reviewed in in Appendix \ref{coherent}. 
 Furthermore, the vacuum energy itself and the density of states  in all these regions  are  also slightly different because the vacuum condensate (\ref{result}) assumes slightly different numerical values in all these regions as it depends on $\mu, T$ and $\theta$.

Precise computations of all these   coherent $\cal{CP}-$ violating QCD effects are hard to carry out explicitly as they require  solution of many body effects in unfriendly environment with non-vanishing $\theta, \mu, T$ when even the phase diagram is not yet known, see Fig.  \ref{fig:phase_diagram}. However, the estimation  of the effect is very simple exercise as it must be proportional to $\theta(t)$ at the moment 
when the domain walls start to form. Numerically, this parameter $|\theta(t)/\theta_0|\sim 10^{-2}-10^{-3}$ could be quite small, and naively may lead only to minor effects $\leq 10^{-2}$. However, the crucial point here is that while this coupling is indeed small on the QCD scale, it nevertheless effectively long-ranged and long lasting, in contrast with conventional random QCD processes. As a result, these coherent  
 $\cal{CP}-$ violating effects will produce large effects of order one as explicit computations carried out below  show.

 The crucial element in making such assessment is the observation that these small coherent changes occur in the entire volume of a  nugget. In other words,   the generation of the chemical potential, accumulation  the phase difference  (which eventually leads to  the accretion of the  baryon charge)   etc are proportional to the  volume of the nugget, 
 \be
 \label{delta_B}
 \Delta B\sim \theta (t) V, ~~~~|\theta(t)/\theta_0|\sim 10^{-2}-10^{-3},
 \ee
 where $\Delta B$ is the baryon charge difference in accumulation between the nuggets and anti-nuggets. This expression should be 
    contrasted with effect due to the local spontaneous violation of the $\cal{C}$-symmetry (as discussed  in  \cite{Liang:2016tqc} and reviewed  in section \ref{spontaneous}) which is proportional to the surface of a nugget, $ B\sim S$, see eq.(\ref{N}). 
     
 While all individual events  are proportional to small parameter $\theta\ll 1 $, and therefore, numerically small,  this difference (between a typical size for  nuggets vs anti-nuggets) eventually will be order of one as a result of long  lasting evolution,  as is shown  in next subsections. This strong effect of order of one has been phenomenologically accounted for by    parameter $c(T)$ in formula (\ref{ratio2}), which automatically leads to    the main   consequence  of this framework expressed by relation (\ref{Omega}).

\subsection{Basic equations}\label{analytical}
We start with     brief  overview of the work   \cite{Liang:2016tqc}, where it was assumed that  the $\cal CP$ symmetry is   globally conserved, i.e.
$\la \theta (t)\ra=0$ when average over entire Universe. In that  paper 
the effective  interaction of the axion domain wall  with Fermi fields was approximated  as follows
\begin{equation}
\label{eq:L4}
\mathcal{L}_4
=\bar{\Psi}(i\slashed{\partial}-me^{i[\theta(x)-\phi(x)]\gamma_5}-\mu\hat{\gamma}_0)\Psi,
\end{equation}
where parameter $m$ should not be interpreted as the quark's mass. Rather, it is much more appropriate to interpret  parameter $m$ as the expectation value of the $\la {\rm det } ~\bar{\psi}_L^f \psi_R^f\ra$ from eq (\ref{result}) which has a typical QCD scale and which will be always generated even in deconfined regime at $T>T_c$. The Fermi field $\Psi$ should be interpreted as the low-energy (soft) component of the original quark fields, while a high energy component has been integrated out to generate parameter $m$ entering (\ref{eq:L4}).   The equation ($\ref{eq:L4}$) is the standard form for the interaction between the pseudo-scalar fields (axion $\theta(x)$ and $\eta'$ field parametrized by $\phi (x)$)  and the fermions (quarks and antiquarks) which respect all relevant symmetries in presence of nonzero chemical potential $\mu$. 

Another key element  from  \cite{Liang:2016tqc} which is relevant for our present studies   is the effective Lagrangian describing the  dynamics of an oscillating domain wall bubble 
\be
\label{lagrangian}
L&=&\frac{4\pi\sigma R^2(t)}{2} \dot{R}^2(t)  - 4\pi\sigma R^2(t) \\
&+&\frac{4\pi R^3(t)}{3}\Delta P(\mu) +\left[{\rm other ~terms}\right]. \nonumber
\ee
The  equation (\ref{lagrangian})   describes the time evolution of the closed spherically   symmetric domain wall bubble of radius $R(t)$.  The pressure term $\Delta P=P^{\rm (Fermi)} +P^{\rm (Bag)}-P^{\rm (out)}$ in eq. (\ref{lagrangian}) can be parametrized  as \cite{Liang:2016tqc}
\begin{equation}
\label{eq:pressure}
\Delta P=\frac{g^{\rm in}T^4}{6\pi^2}I_4(b)-E_B\theta(b-b_1)\left(1-\frac{b_{1}^2}{b^2}\right)-\frac{\pi^2g^{\rm out}T^4}{90}, \nonumber
\end{equation}
where $g^{\rm in}\simeq2N_{c}N_{f}$ and $g^{\rm out}\simeq\left(\frac{7}{8}4N_{c}N_{f}+2(N_{c}^{2}-1)\right)$ are degeneracy factors for inside and outside of the bubble respectively; $b=|\mu|/T$ is the dimensionless parameter to    be used frequently in what follows; $I_4(b)$ is a specific  type of Fermi integrals, see Appendix \ref{Appendix:Fermi_integrals}; finally $E_B\sim(150~{\rm MeV})^4$ is the famous ``bag constant'' from the MIT bag model,  
which turns on in the hadronic phase at small $\mu< \mu_1$, while it vanishes  in CS phase at large $\mu> \mu_1$.  

The evolution of the bubble is governed by the   following differential equations \cite{Liang:2016tqc}
 \begin{subequations}
\label{eq:evolution}
\begin{equation}
\label{eq:evolution_B}
\frac{d}{dt}B_{\rm wall}=0,~~~~ B_{\rm wall} =  \frac{g^{\rm in} ST^2}{2\pi}I_2(b)
\end{equation}
\begin{equation}
\label{eq:evolution_R}
\sigma\ddot{R}=-\frac{2\sigma}{R}-\frac{\sigma\dot{R}^2}{R}
+\Delta P(b)-4\eta\frac{\dot{R}}{R}.
\end{equation}
\end{subequations}
where $\sigma\simeq9f_a^2m_a$ is the domain wall tension, while $\eta \simeq m_\pi^3$ is the QCD viscosity. The baryonic charge for  nuggets   can be expressed in terms of the Fermi integral $I_2(b)$  given in Appendix  \ref{Appendix:Fermi_integrals}.  
Eq.  (\ref{eq:evolution_B}) describes an implicit dependence of chemical potential on size of the nugget, $\mu(R)$.
It should be substituted to eq. (\ref{eq:evolution_R})  to arrive to the differential equation which describes the time evolution of the nugget $R(t)$. The corresponding analysis has been performed    in  \cite{Liang:2016tqc} with the basic result that 
the solution can be well approximated by slow damping oscillations with typical frequency $m_a$ around $R_{\rm form}$ as presented by eq. (\ref{eq:6.R2}).

As we already mentioned, in the preceding studies \cite{Liang:2016tqc} on the bubble evolution we had assumed that  the  $\cal CP$-symmetry is  conserved as no interaction with  the coherent axion field $\theta(t)$ was included  into the consideration. Now we wish to  include the coherent axion field $\theta(t)$, 
given by eq.(\ref{axion})  into the equations. 
The most important change which   occurs as a result of the interaction of the Fermi fields with the background axion field is the accumulation of the phase  (\ref{phase}) as a result of the coupling $\Psi$ field   with   axion field $\theta(t)$, which can be interpreted as the coherent Berry's phase,  as explained in Appendix \ref{coherent}. 

 To proceed with the computations one could follow the procedure developed in \cite{Liang:2016tqc}, rewrite the Fermi fields in 2d notations (accounting for the domain wall background) and represent the extra term to the effective Lagrangian due to the 
coupling with external axion field as follows  

\be
\label{eq:A1}
\Delta \mathcal{S}&=&-\frac{1}{\pi}\int dz\mu (z)\partial_z\theta (t), \\
A&\equiv& -\frac{\delta\Delta \mathcal{S}}{\delta \mu}
=\frac{1}{\pi}\int_{R_{\rm in}}^{R_{\rm out}}dz\cdot \left(\frac{\partial \theta}{\partial t}\right)\frac{1}{\dot{R}},\nonumber
\ee
where the end points of the integral $z= (R_{\rm in}, R_{\rm out})$ should be interpreted as the typical distances  describing the regions inside and outside the   bubble of  a typical size $z\simeq   R$. 

Few comments are in order. 
First, one can explicitly see from eq. (\ref{eq:A1}) that the effect is different for nuggets and anti-nuggets
as the sign of $\mu$ is opposite in these cases, while the background field $\theta(t)$ remains  the same in entire visible Universe.   Secondly,  parameter $A$ can be interpreted as an additional baryon charge (per degree of freedom) accumulated by a nugget during a single nugget's cycle. This coefficient can be approximated as follows 
\begin{equation}
\label{eq:A2}
A\simeq\frac{\Delta\theta}{\pi},
\end{equation}
where $\Delta\theta$ can be interpreted as the variation of the axion field during a single-nugget's    cycle.
 One should emphasize that 
 the correction (\ref{eq:A2})   is not the only source leading to the disparity  between nuggets and anti-nuggets. In fact many other $\cal{CP}$ violating processes  may also contribute to the differences in accumulating baryon charges for nuggets and anti nuggets. We expect that all these effects mentioned in footnote \ref{processes} are equally important  as they must be   proportional to the $\cal{CP}$ violating parameter $\theta(t)$. Therefore,  the corresponding effects  may be effectively accounted for by a modification  of the numerical coefficient  $A$ entering (\ref{eq:A2}). As we argue below, the outcome of the calculations  is not very sensitive to a specific value of coefficient $A$. 
 Therefore, neglecting a large number of different $\cal{CP}$ violating processes  mentioned in footnote \ref{processes} (which can be accounted for by modifying   the numerical coefficient $A$) does not affect   the main result of our analysis. 
 
  Our next remark is the observation that the phase (\ref{axion}) and corresponding extra energy  (\ref{eq:A1})  is accumulated coherently by large portion of  quarks inside  of a  nugget of volume $V$, while corresponding correction to the vacuum energy  obviously vanishes outside the nuggets where $\mu\simeq 0$. To compute the correction to the baryon charge of a nugget   due the coupling to the axion $\theta$ field one should multiply (\ref{eq:A2}) to the degeneracy factor, i.e. 
 \be
\label{eq:Btheta}
B_\theta^{(\pm)}
&=&\pm g^{\rm in} AV\int\frac{d^3k}{(2\pi)^3}\frac{1}{\exp(\frac{k}{T}-b)+1}\\
&=&\pm\left(\frac{g^{\rm in} ST^2}{2\pi} I^{(\pm)}_2(b)\right)\cdot\left(\frac{ART}{3\pi}\frac{I^{(\pm)}_3(b)}{I^{(\pm)}_2(b)}\right), \nonumber
\ee
where we assume that the chemical equilibration is sufficiently fast such that one can use one and the same $\mu$ 
within entire volume of the nugget. Furthermore, in writing down equation (\ref{eq:Btheta}) we assumed that the majority of  quarks in volume $V$ move coherently during  the bubble oscillations. In fact, the coherence might not be so perfect, and only some portion of the quarks inside the nuggets might move coherently as a macroscopical system. It is very hard to estimate the corresponding suppression factor within our framework formulated in terms of a single macroscopical variable  $R(t)$. The corresponding suppression factor may be effectively accounted for by some modification   of the numerical coefficient $A$ entering   (\ref{eq:Btheta}).  As we already previously mentioned our results are not very sensitive to an absolute value of numerical value of $A$ as long as it is not exceedingly small.  Therefore, we assume that all correction factors mentioned previously in footnote \ref{processes} and suppression factor due to not perfect coherent motion are implicitly included in coefficient $A$ in analysis which  follows.

 Our next remark goes as follows. In eq. (\ref{eq:Btheta}) we presented the expression for $B^{(\pm)}_\theta$ as a combination of two factors. The first factor is precisely the expression for the baryon charge $B_{\rm wall}$ from eq. (\ref{eq:evolution_B}) generated spontaneously as described in \cite{Liang:2016tqc}. The second factor represents the correction to the accumulated baryon  charge due to the coupling of the $\cal CP$ odd axion field with quarks inside the nugget.  It is important to emphasize that it is numerically suppressed due to small
$\cal CP$ violating phase (\ref{axion}) parametrized by small numerical coefficient $A$. However, it is strongly enhanced by
large numerical factor  proportional to the size of the system.    
In contrast to the baryon accumulation in Eq.  (\ref{eq:evolution_B}) which is proportional  to the surface, the $ B_\theta^{(\pm)} $ contribution  is proportional to the volume of the system (\ref{eq:Btheta}), in agreement with qualitative arguments (\ref{delta_B}) of the previous section. 

Our next comment  is as follows. 
Naively, one could think  that the accumulation of  the baryon charge by  a nugget due to the coherent interaction with the axion field $\theta$  will be very  fast since it is proportional to the volume $V$ of the nugget according to (\ref{eq:Btheta}).  However, the accumulation is  in fact  quite  slow. The point is that     the axion $\theta$ field   oscillates with time,    $\sim\cos(m_at)$ and  the accumulated baryon charge   is  almost washed out  during a   complete cycle of the axion $\theta(t)$ field. Nevertheless, the cancellation is not quite complete because  the axion field slowly reduces its    amplitude. The main reason for this amplitude's decay is the emission of real physical axions (due to misalignment mechanism) which is the source of some non- equilibrium dynamics.   Precisely this slow decrease of the axion amplitude   leads to a non-vanishing  $\Delta \theta$ entering (\ref{eq:A2}), and eventually generates  the disparity   between nuggets and anti-nuggets. 

 Recapitulate: a tiny portion of the accumulated baryon charge remains in the nuggets after every single complete cycle  such that the disparity between nuggets and anti-nuggets will be accumulated but at very slow pace as a result of very large number of oscillations. Based on this fact, we can therefore assume $A$ as an adiabatic ``constant''  for each given cycle in the  analysis in  following subsections \ref{qualitative} and \ref{slow}.

 Our   final remark regarding  eq.  (\ref{eq:Btheta}) is as follows. This formula was derived  by considering the coherently moving  quarks of  the nuggets  in the background of the axion field (\ref{axion}). This approximation is only justified as long as the effect due to the background field is sufficiently small.
 Formally, the effect (\ref{eq:Btheta}) due to the axion field  must be much smaller than the initial accumulation of the baryon charge   (\ref{eq:evolution_B}) due to the spontaneous local  symmetry breaking when the original chemical potential $\mu$ is generated. 
 Such a condition must be imposed on our system to avoid any complications related to  accounting for  the feedback (back reaction) of the coherent Fermi field on the background field. We satisfy this condition  by requiring that the expression in the second bracket in  (\ref{eq:Btheta}) is (marginally) smaller than unity, i.e.
 \begin{equation}
\label{eq:A_condition}
\frac{B^{(\pm)}_\theta}{B_{\rm wall}}\sim\left(\frac{ART}{3\pi}\frac{I^{(\pm)}_3(b)}{I^{(\pm)}_2(b)}\right)\lesssim 1.
\end{equation}
This condition can be also understood as the requirement that the accreted baryon charge $B^{(\pm)}_\theta$ due to the background field does not change the original boundary conditions imposed on the quark's fields in the domain wall background.  Precisely these boundary conditions  determine the sign of the $B_{\rm wall}$ as a result of the  local spontaneous symmetry breaking effect as discussed in \cite{Liang:2016tqc}. 

Furthermore, 
precisely these boundary conditions generate  the initial    chemical potential of a nugget which enters (\ref{eq:A1}). 
The requirement (\ref{eq:A_condition}) states that the influence of $B^{(\pm)}_\theta$ on the initial  axion domain wall background should be sufficiently small. Although condition (\ref{eq:A_condition}) is exact, it is not technically  useful  to implement  in the following analysis because it contains complicated functions of $R$ and $\mu$. To get rid of such cumbersome dependence, we use the fact $I_3(b)/I_2(b)$ can be approximated for small $\mu$ with sufficiently  high accuracy as $I_3(b)/I_2(b)\simeq3/2$ at  $b=|\mu|/T\ll 1$, see  Appendix \ref{Appendix:Fermi_integrals}. With this approximation we  obtain a much  more  useful, practical  and transparent condition
\begin{equation}
\label{eq:a}
a (t)\equiv\frac{AR_0T}{2\pi}\simeq\left|\frac{B^{(\pm)}_\theta}{B_{\rm wall}}\right|, ~~~~
|a(t)|\lesssim1.
\end{equation}
Note that we define this simplified condition in terms of a single    parameter $a(t)$. By monitoring the value of $a(t)\lesssim1$, we can precisely determine  whether our analysis is valid and justified or it is   about to break down. In what follows, we will express formula in terms of $a(t)$ up to the linear order since we are only interested in region of sufficiently small $a\lesssim1$ where our analysis is marginally justified. It is quite obvious that $a(t)$ can be treated as adiabatic parameter slowly changing with time as the typical time scale for variation $a(t)$ is determined by changing of the axion field during a single cycle (\ref{eq:A2}).

After these comments   we  can now follow the     procedure developed in   \cite{Liang:2016tqc} to account for the $\cal CP$ violating effects by replacing eq.(\ref{eq:evolution_B}) by  its generalized version in  the form
\begin{equation}
\label{eq:evolution_B2}
\frac{d}{dt}\left(B_{\rm wall}+B^{(\pm)}_{\rm \theta}\right)=0. 
\end{equation}
Similar to  \cite{Liang:2016tqc} we treat eq. (\ref{eq:evolution_B2}) as an implicit relation between $\mu(t) $ and $R(t)$ which should be substituted into eq. (\ref{eq:evolution_R}) to arrive to a single differential equation which governs the dynamics of $R(t)$ as a function of time $t$
\begin{equation}
\label{eq:R_simp2}
\sigma \ddot{R}(t)
=-\frac{2\sigma}{R}-\frac{\sigma\dot{R}^2}{R}
+\Delta P^{\pm}[R]-4\eta\frac{\dot{R}}{R},
\end{equation}
where the pressure term is now a   function of bubble radius $R(t)$, rather than $\mu$. It can be approximated   as follows
\be 
\label{eq:P_simp}
 &&\Delta P^{(\pm)}[R]
\simeq\frac{g^{\rm in}T^4}{6\pi^2}\left[2\pi f^{(\pm)}(R)+\left(f^{(\pm)}(R)\right)^2\right]  \nonumber \\
&&-\frac{\pi^2g^{\rm out}}{90}T^4-E_B\theta(b-b_1)\left(1-\frac{b_{1}^2}{b^2}\right).
\ee
 where functions $f^{(\pm)}(R)$ entering (\ref{eq:P_simp}) have  been computed  in Appendix \ref{technics} and can be approximated  as follows
\be 
\label{f}
f^{(\pm)}(R) \simeq \frac{\pi^2}{12}\frac{T_{0}^2R_0^2}{T^2 R^2}
\left[1\mp a\left(\frac{4}{9}\frac{T_{0}}{T}\sqrt{\frac{\pi^2}{12}}
+\frac{R}{R_0}\right)\right].  ~~~
\ee
The new element in comparison with our previous studies  \cite{Liang:2016tqc} is that the pressure $\Delta P^{(\pm)}[R]$ is now different for nuggets and 
anti-nuggets because the functions $f^{(\pm)}(R)$ which determine their dynamics are quite distinct.  The corresponding difference is determined by the coefficient 
(\ref{eq:a}) which is explicitly proportional to the $\cal{CP}$-odd  parameter $A$ as defined by eqs. (\ref{eq:A1}), (\ref{eq:A2}). Precisely this 
difference, as we discuss below, determines the imbalance in evolution of the nuggets and anti-nuggets. 

 In what follows, we keep the temperature $T$ to be constant, and furthermore, we shall  assume $T_0/T\simeq1$ to simplify our numerical analysis. 
 The justification for the first assumption (the temperature $T$ is kept  constant)  is that all relevant processes (including the nugget's oscillations)  have the time scales which are much shorter    than a typical cosmological time scale  when temperature $T$ and the axion mass $m_a(T)$ slowly   vary. 
 
   Indeed, as  the temperature scales with cosmic time as $T\sim t^{-1/2}$ the corresponding variations of the temperature during a single axion oscillation, determined by time $\Delta t\sim m_a^{-1}$, is very tiny, $\Delta T/T\sim \Delta t/t\sim (m_a t)^{-1}$. Numerically, it represents extremely small correction $\Delta T/T\sim 10^{-5}$  for $m_a\sim 10^{-6}$eV as the cosmic time $t$ corresponding to the temperature  $T_0$ is of order  $t\sim 10^{-4}$s. It is known that the axion mass $m_a(T)$ experiences very sharp changes with the temperature   $m_a(T)\sim T^{-n}$ with exponent $n\sim 8$, see \cite{Kitano:2015fla,Bonati:2015vqz,Borsanyi:2016ksw,Petreczky:2016vrs}.
Nevertheless,    the axion mass    does not vary much during a single axion oscillation.  Indeed, during   time $\Delta t\sim m_a^{-1}$ the axion mass receives very tiny correction, $\Delta m_a/m_a\sim n(\Delta T/T)\sim n (m_a t)^{-1}\ll 1$. 

Furthermore, one can argue that the nuggets make very large number of oscillations during few cycles of the axion field $\theta$ when    the axion mass $m_a (T) $ and the temperature $T$ experience very insignificant  relative corrections. This justifies our assumption that $m_a(T)$ and $T$ can be kept constant in our studies of the nugget's dynamics. We refer to Appendix 
\ref{Appendix:CPodd_numerical}  with corresponding estimates and details.

 One can rephrase these arguments  in slightly different way as follows.  Slow change of the temperature $T$ and the axion mass $m_a(T)$ is accompanied by slow variation of the axion amplitude $\theta(t)$. Precisely this $\theta$ variation  eventually  leads to the slow accumulation of the disparity between nuggets and anti-nuggets as argued below. The effects of variation of the dynamical equations such as (\ref{eq:R_simp2}) due to tiny changes of the temperature and the axion mass $\Delta T, \Delta m_a$ do not affect our analysis because the main source of the disparity  is explicitly proportional to the chemical potential $\mu$ according to eq. (\ref{eq:A1}) while the small changes of  temperature and the axion mass $\Delta T, \Delta m_a$  during the evolution  contribute equally to both types of species and play the sub-leading role. 
  
  Another  assumption ($T_0/T\simeq1$)  represents a pure technical simplification in   our  analysis to demonstrate that the disparity  in evolution for   different species is order of one effect. 
 In fact, one can argue that this imbalance (between the nuggets and anti-nuggets)   becomes even more pronounced 
 if  one accounts the decrease  of temperature  with time.

  In principle, the equation (\ref{eq:R_simp2}) can be solved numerically (without a large number of simplifications and assumptions mentioned above)
 which would determine the  dependence of $R(t)$ and $\mu(t)$ on  time. 
  The corresponding results of these   numerical studies   are presented in Appendix \ref{Appendix:CPodd_numerical}. 
  These numerical results are fully consistent with our analytical (simplified) treatment of the problem, which is   the subject of the following  subsections
  \ref{qualitative} and \ref{slow}.

\subsection{Time evolution.   }\label{qualitative}  
First, we find  the equilibrium condition when the ``potential" energy in eq.(\ref{eq:R_simp2}) assumes its minimal value, similar to the procedure 
  carried out in  \cite{Liang:2016tqc}. The corresponding minimum condition 
is determined by equation 
\begin{equation}
\label{eq:6.Rform}
\frac{2\sigma}{R_{\rm form}}=\Delta P^{(\pm)}[R_{\rm form}], 
 \end{equation}
 where  $\Delta P^{(\pm)} $ is defined  by eq.(\ref{eq:P_simp}).  The difference with the previously  studied case   \cite{Liang:2016tqc} is that 
 $\Delta P^{(\pm)} $ is now  different for nuggets and anti-nuggets. Therefore the equilibrium solution 
 $R^{(\pm)}_{\rm form}$ will be also different for two  different species.  This is the key point of the present studies. 
 Another distinct feature  is that $R^{(\pm)}_{\rm form}$   slowly  varies  with time because the axion background adiabatically  changes with time.
 The corresponding variation explicitly enters equation (\ref{eq:evolution_B2}) and implicitly equation (\ref{eq:R_simp2}).
  
 As the next step we want to see how these different equilibrium solutions are approached when time evolves. 
We follow the conventional technique and  expand (\ref{eq:R_simp2})  around the equilibrium values $R^{(\pm)}_{\rm form}$ to arrive to an equation for a  simple damping oscillator:
\begin{equation}
\label{eq:6.R1}
\frac{d^2\delta R^{(\pm)}}{dt^2}+\frac{2}{\tau^{(\pm)}}\frac{d\delta R^{(\pm)}}{dt}+(\omega^{(\pm)})^2\delta R^{(\pm)}=0, ~~~ 
\end{equation}
where $\delta R^{(\pm)}\equiv [R(t)-R^{(\pm)}_{\rm form}]$ describes the deviation from the equilibrium position, while new parameters   $\tau^{(\pm)}$ and $\omega^{(\pm)}$ 
describe the effective  damping coefficient  and   frequency of the oscillations. Both new coefficients   are    expressed in terms of  the original parameters entering (\ref{eq:R_simp2}) and are given by  
\begin{subequations}
\label{eq:6.R1tw}
\begin{equation}
\label{eq:6.R1t}
\tau^{(\pm)}=\frac{\sigma}{2\eta}R^{(\pm)}_{\rm form}
\end{equation}
\begin{equation}
\label{eq:6.R1w}
(\omega^{(\pm)})^2=\left.-\frac{1}{\sigma}\frac{d\Delta P^{(\pm)}  (R)}{dR}\right|_{R^{(\pm)}_{\rm form}} -\frac{2}{[R^{(\pm)}_{\rm form}]^2}.
\end{equation}
\end{subequations}
The expansion (\ref{eq:6.R1}) is justified, of course, only  for small oscillations about the minimum determined by eq.(\ref{eq:6.Rform}), while the oscillations determined by original equation (\ref{eq:R_simp2})  are obviously  not small.  
However, our  simple analytical treatment  (\ref{eq:6.R1}) is quite instructive and gives a good qualitative understanding of the system.
The difference with previously  studied case    \cite{Liang:2016tqc} is of course the emergence of two different solutions for two different species characterized by parameters (\ref{eq:6.R1tw})   which also slowly vary with  time. 
Our numerical studies presented in Appendix \ref{Appendix:CPodd_numerical}  fully support the qualitative picture presented in this subsection.

  The most important conclusion of these   studies is that nuggets and anti nuggets oscillate in very much the same way 
  as we observed in our previous studies  and well approximated  by eq. (\ref{eq:6.R2}). However, their evolution proceed in somewhat different manner 
  now    as a result of $\cal{CP}$ violating terms as discussed above. 
  
  To analyze  this difference in a quantitative way we introduce  parameter  $\Delta R(t) \equiv |R^+(t)-R^-(t)| $ which measures this difference
  between nuggets'  and anti nuggets' sizes  assuming that the same initial conditions are imposed, i.e. $R^{\pm}_0=R_0\sim m_a^{-1}$. The parameter  $\Delta R(t)$ is the most important quantitative characteristic  for our present studies as it shows how the sizes of nuggets and anti-nuggets evolve with time.  We shall  demonstrate that $\Delta R /R$ becomes  order of one  when the condition 
  (\ref{eq:a}) is still marginally satisfied.  Our simplified computations (by neglecting the back reaction which modifies the background) obviously break down  when the ratio  (\ref{eq:a}) approaches one. However, once a sufficiently large effect is generated we do not expect that it  can be completely washed out by further evolution. Rather, it is expected that once  a  large effect 
  is generated it remains to be sufficiently   large effect of order one, though a precise numerical coefficient might be different from  our qualitative analysis.   In fact,   precise magnitude of $\Delta R/R$  is very hard to compute as there are many other effects mentioned in footnote \ref{processes} 
  which influence the time evolution and equally contribute to $\Delta R/R$. 
  
    Therefore, the key result of our analysis   is that  $\Delta R (t)$   fluctuates with time, but always approaches a non-vanishing magnitude  of order of one 
  during the long cosmological evolution.  The fluctuations for $R^{\pm}(t)$ are well approximated by  eq.(\ref{eq:6.R2})  where key parameters (\ref{eq:6.R1tw}) now are different for different species. Numerical results presented on Fig. \ref{fig:numerical} support this qualitative analysis presented above.

\subsection{ Disparity in sizes for nuggets and anti-nuggets.  }\label{slow}
While numerical results presented on Fig.  \ref{fig:numerical} explicitly show that the difference  between typical sizes of  nuggets and anti-nuggets
becomes of  order of one effect during the time evolution,
we would like to understand this important feature of the system using    analytical, rather than numerical arguments. This subsection is devoted precisely to such an  analysis.  

In what follows, we would like to estimate $\Delta R(t) \equiv |R^+(t)-R^-(t)|$    at $t\rightarrow \infty$  when the system approaches its equilibrium, i.e 
we are interested in the difference   $\Delta R_{\rm form}\equiv |R^+_{\rm form}-R^-_{\rm form}|$ between nuggets and anti-nuggets. The equilibrium values for each species can be approximated as 
\begin{equation}
\label{eq:5C Rform}
\begin{aligned}
\frac{2\sigma}{R_{\rm form}}
\simeq \frac{g^{\rm in}T^4}{6\pi^2}
\left[f^{(\pm)}(R_{\rm form})\right]^2
\end{aligned}
\end{equation}
where we simplified the  original equations eqs. (\ref{eq:P_simp}) and (\ref{eq:6.Rform})  
by  keeping  the numerically dominant terms in region when a typical radius of a nugget is considerably dropped from its initial value, i.e. $R_{\rm form}\lesssim0.5 R_0$, see  in Appendix \ref{Appendix:CPodd_numerical} for the details. For  further simplifications and for illustrative purposes  we keep 
  only  the leading  terms $\sim a$, similar to eq. (\ref{f}). Precisely these terms     eventually lead to the disparity between nuggets and anti-nuggets, as we already discussed. With these  simplifications  the formation radius for nuggets and anti-nuggets  can be approximated as follows 
\be
\label{eq:5C Rform1}
R^{\pm}_{\rm form}
&\simeq \la R_{\rm form}\ra
\cdot\left[1\mp\frac{2a_c}{3}
\left(\frac{\la R_{\rm form}\ra}{R_0}+\frac{2\pi}{9\sqrt{3}}\right)
\right],  
\ee
  where $\la R_{\rm form}\ra$ is defined as the average size   of different species,
\be 
\label{eq:5C Rform2}
\langle R_{\rm form}\rangle
&\equiv \frac{1}{2}\left|R^+_{\rm form}+R^-_{\rm form}\right|
\simeq R_0\left[\frac{\pi^2 g^{\rm in}T^4}{12^3}
\frac{R_0}{\sigma (T)}\right]^{1/3}, ~~~~
\ee
while $a_c$ is defined as the critical value at so-called decoherence time, $t_{\rm dec}$ 
when the   axion field is loosing  its coherence\footnote{\label{coherence}The decoherence time, $t_{\rm dec}$ is determined by a number of different processes, including the time scale for the axion field to decay  into the randomly distributed DM axions,  see footnote 13 in \cite{Liang:2016tqc} for comments on this matter.} 
on the scale of the Universe, such that the disparity between nuggets and anti-nuggets does not further evolve after $t_{\rm dec}$, 
\be
\label{a_c}
a_c\equiv a(t\rightarrow t_{\rm dec}), ~~~ a_c < 1.
\ee
In eq. (\ref{a_c}) we also assumed that  $a_c < 1$ is sufficiently small when  the condition (\ref{eq:a}) is marginally satisfied. 

Few comments regarding (\ref{eq:5C Rform1}) and  (\ref{eq:5C Rform2}) are in order. First of all, 
as we already mentioned, we kept  the temperature $T$ to be  constant in our computations. 
We can now treat  $T$ in (\ref{eq:5C Rform2}) as an adiabatic parameter which slowly decreases with time as the temperature slowly  approaches the QCD transition temperature $T_c$ from above. During this evolution the domain wall tension $\sigma (T)$ approaches its final value $\sigma\rightarrow 9f_a^2m_a$ at the QCD transition point $T=T_c$ where the chiral condensate forms. This slow change of the formation radius $\la R_{\rm form}\ra$ is perfectly consistent with the numerical results presented on Fig.  \ref{fig:numerical}.  

Secondly,  using equation (\ref{eq:5C Rform1})  and our estimate (\ref{eq:5C Rform2}) for $\la R_{\rm form}\ra\simeq 0.6R_0$ one can approximate  the disparity in   sizes $\Delta R_{\rm form}$ between two different species  
   as follows
 \begin{equation}
\label{eq:5C Rform_ratio}
\frac{\Delta R_{\rm form}}{\la R_{\rm form}\ra}
\simeq  ~\frac{4}{3} \cdot  a_c. 
\end{equation}
 This estimation suggests that the difference  in baryon charges between the nuggets and anti-nuggets can be approximated as follows
 \begin{equation}
\label{eq:V_difference}
\frac{\Delta B}{\la B \ra}\simeq \left(\frac{\Delta R}{\la R_{\rm form}\ra}\right)^3
\simeq3\frac{\Delta R}{\la R_{\rm form}\ra}
\simeq 4a_c.
\end{equation}
   In other words,   even for relatively  small $a_c\simeq (0.1-0.2)$ where our background approximation remains marginally valid according to condition (\ref{eq:a}),  the disparity in baryon charges (\ref{eq:V_difference}) between nuggets and anti-nuggets could be quite large and can easily satisfy the basic assumption (\ref{ratio2}) with $|c(T)|\sim 1$.     

Our last comment is   to elaborate on  the physical meaning of parameter $a(t)$ which is defined by (\ref{eq:a}). This parameter  enters (\ref{eq:V_difference}) and plays an important  role in our discussions and estimates which follow. The phenomenological parameter $a(t)$ has been  introduced into our analysis  to describe very small variation (\ref{eq:A2}) of the axion field after each single cycle during the nuggets' evolution. If the axion field were perfectly periodic with the original amplitude $\theta_0$ being kept  constant, then our parameter $a(t)$ would be identically zero as every consequent  cycle of  the axion field would wash out the asymmetry it produces during a previous cycle, 
as we already mentioned at the end of subsection \ref{analytical}. However, the axion coherent field decays as a result of the production of the real propagating axions (as well as result of  many other processes mentioned in footnote \ref{processes}). We parametrize these changes of the system  by function  $a(t)$ assuming that initially $a(t=0)=0$ vanishes, but slowly increases its value during the nugget's evolution. It reaches the maximum value
$a_c$ when the largest possible disparity between nuggets and anti-nuggets is achieved\footnote{\label{a_large}In our numerical studies, we also  assume that $a(t)$    never becomes  too big   which would violate our approximations, and would  change the boundary condition imposed by the sign 
of the chemical potential $\mu$  as expressed by eq.(\ref{eq:A_condition}). We shall make few technical comments  in next section \ref{survival} on how to proceed with computations if parameter $a(t)$ becomes numerically large and   our treatment of  $a(t)$ as a small  correction is obviously breaks down.}  as determined by eq. (\ref{eq:V_difference}). 
The disparity (\ref{eq:V_difference}) cannot be washed out  after $t_{\rm dec}$, as we discuss in the next section \ref{difference}, because the   axion field is lost its coherence on the scale of the Universe, and cannot   wash out the previously generated imbalance (\ref{eq:V_difference}).

We  conclude this section with the following remark. The main goal of this section was to demonstrate that even relatively small $\cal CP$-violating   coupling (\ref{eq:A1}), (\ref{eq:A2}) of the coherent axion field with quarks (anti-quarks) inside the nuggets (anti-nuggets) generically produces a large effect of order one as a result 
of a coherent long lasting influence of the axion field on the dynamics of the nuggets. We presented some semi-analytical estimates  expressed by eq. (\ref{eq:V_difference}) supporting this claim. The 
numerical results, obtained without a large number of simplifications and presented  in Appendix \ref{Appendix:CPodd_numerical} (see  specifically 
Fig.\ref{fig:numerical}) also reinforce  the analytical  analysis of this subsection. 

The significance of the result (\ref{eq:V_difference}) is that the disparity between nuggets and anti-nuggets   unambiguously implies that our main assumption formulated 
  as eq.(\ref{ratio2}) is strongly supported by the computations of this section. Needless to say that eq.(\ref{ratio2}) 
  is essentially equivalent to our  generic fundamental consequence   (\ref{Omega}) of this framework suggesting that the visible and dark matter densities are  of the same order of magnitude $\Omega_{\rm dark} \approx \Omega_{\rm visible}$ irrespectively to the parameters of the system.

\section{Robustness of the imbalance  between nuggets and anti-nuggets}\label{difference}

In this  section we would like to argue that the results of previous section \ref{sec:evolution} are  very robust in a sense that they are   not very sensitive 
  to the fundamental parameters of the system such as the axion mass $m_a$ or initial misalignment angle $\theta_0$. In particular, we shall argue in section \ref{sect:interpretation} that the results of previous section  are also insensitive     to a large number of technical simplifications we have made in previous section.   Furthermore, we also present some  arguments in section  \ref{survival} that  the  imbalance  (\ref{eq:V_difference}) survives 
the subsequent evolution of the system at $t>t_{\rm dec}$ after   the axion field loses its coherence.

\subsection{Insensitivity to the axion mass $m_a$ and  initial misalignment angle $\theta_0$}\label{sect:interpretation}
 First of all, we want to argue that the disparity between nuggets and anti-nuggets expressed by eq. (\ref{eq:V_difference}) is not very sensitive to 
the axions mass $m_a$ which itself varies with the temperature in the range $T_a\leq T\leq T_c$ and approaches  its final value at the QCD transition at $T_c$, as explained in section \ref{formation}. As a result of this ``insensitivity" of eq.  (\ref{eq:V_difference}) 
to $m_a$, the main 
consequence  of the entire framework expressed as  eqs. (\ref{Omega}) and  (\ref{ratio2})   is also insensitive to the axion mass.

 The main reason for this claim is that we are interested in relative ratio (\ref{eq:V_difference}) between nuggets and anti-nuggets rather than in absolute value $\la R_{\rm form}\ra $ of a nugget  which is obviously sensitive to the axion mass as it scales as $\la R_{\rm form}\ra \sim m_a^{-1}$ according to  (\ref{eq:5C Rform2}).
  The ratio (\ref{eq:V_difference}) on other hand 
 is not very sensitive to the absolute value of the axion mass. This feature is manifestly seen    on Fig.  \ref{fig:numerical} where two plots for $m_a=10^{-4} {\rm eV}$ and $m_a=10^{-6} {\rm eV}$ are almost identically the same. 
 
 To summarize the argument regarding the axion mass: the size of a nugget  and its total baryon charge are highly  sensitive to axion mass $m_a$. However, their relative ratio (\ref{eq:V_difference}), which is equivalent to the basic relation (\ref{ratio2}),  is always of order of one and largely $m_a$-independent.  This basic feature  eventually leads to fundamental prediction of this framework,  $\Omega_{\rm dark} \approx \Omega_{\rm visible}$  which is insensitive to the axion  mass $m_a$, in contrast with conventional mechanisms of the axion production when $\Omega_{\rm dark} \sim m_a^{-7/6}$, see  recent reviews   
  \cite{vanBibber:2006rb, Asztalos:2006kz,Sikivie:2008,Raffelt:2006cw,Sikivie:2009fv,Rosenberg:2015kxa,Graham:2015ouw,Ringwald2016}.  
   
Consequentially, we want to argue that the disparity between nuggets and anti-nuggets expressed by eq. (\ref{eq:V_difference}) is not very sensitive to   the  initial conditions of the misalignment angle $\theta_0$.  
This is  because  the disparity (\ref{eq:V_difference})  is determined by 
parameter  $A\simeq\Delta\theta/\pi$ from  Eq. (\ref{eq:A2}) which has a meaning of the accumulated changes during a single cycle. 
The total accumulation of the changes  during a large number of cycles is  determined by  the   relation  (\ref{eq:a}) which slowly approaches some constant value $a_c$ irrespectively what  the original   misalignment angle $\theta_0$ at the initial time was. 

In other words, if the initial  $\theta_0$ was quite small then  it might take a longer period of time before the coefficient $a(t)$ 
assumes its final value $a_c$.  If the initial $\theta_0$ was sufficiently large,   it might take slightly shorter period of time to get to the point when $a(t)$ approaches its finite value (\ref{a_c}). However, in all cases the coefficient $a(t)$   approaches the constant value (\ref{a_c}) of order one 
due to the parametrically  enhanced  factor  $\sim R_0T$ in eq.(\ref{eq:a}) such that even  very tiny initial magnitude of $a(t)$ generates  order of one effect   due to  the long lasting coherent axion field as explained in the text after eq. (\ref{eq:V_difference}).  This behaviour should be  contrasted with conventional mechanisms of the axion production when $\Omega_{\rm dark} \sim \theta_0^2$ is highly sensitive to initial misalignment angle $\theta_0$,  see  recent reviews   
  \cite{vanBibber:2006rb, Asztalos:2006kz,Sikivie:2008,Raffelt:2006cw,Sikivie:2009fv,Rosenberg:2015kxa,Graham:2015ouw,Ringwald2016}. 
  
   The same argument also applies to the initial $T_0$. To be more specific: the  final result for the disparity  is not very sensitive to the initial temperature $T_0$ as the relative imbalance  (\ref{eq:V_difference})  is essentially determined by the final value $a_c$ rather than by some  dynamical features of the system\footnote{To avoid confusion, one should comment here that $a_c (T)$ itself {\it implicitly} depends on the temperature as the axion dynamics is highly sensitive to the temperature. Our claim on ``independence" on $T_0$ refers to {\it explicit} independence of the disparity  (\ref{eq:V_difference}) on $T_0$. Implicit dependence of  (\ref{eq:V_difference})  on temperature through $a_c (T)$ always remains.}. 
   
   Another important dimensional parameter of the system is viscosity $\eta$ which, in particularly,  enters eq. (\ref{eq:6.R1tw}) and  determines the frequency of oscillations, $ \omega^{(\pm)}$ and effective  damping coefficient $\tau^{(\pm)}$ during the evolution. The sizes of the nuggets $R^{\pm}(t)$ are obviously very sensitive to these parameters $ \omega^{(\pm)}, \tau^{(\pm)}$, and therefore, to viscosity $\eta$.
   However, as in our previous discussions,  the    disparity (\ref{eq:V_difference}), which is dimensionless parameter measuring the relative 
   sizes on different species   is not sensitive to this parameter when computed at the very end of the evolution. In other words, it might take 
   longer (or shorter) period of time for different values of $\eta$ to get to  the final  destination determined by eq.  (\ref{eq:V_difference}).
   However, the numerical value  of the disparity    (\ref{eq:V_difference}) always  remains the same and determined by parameter $a_c$ when the 
   axion (initially coherent)  field is lost its conference. 
   To demonstrate this feature we present the behaviour of the system for two different values of the viscosity $\eta$ on   Fig.\ref{fig:different_eta}. As one can see from the plot, the results are   identically the same at the end of the evolution. Formally, it is due to the fact that the viscosity enters the equation with
   $\dot{R}$, and therefore it is not really a surprise that the dependence on $\eta$ diminishes when the system approaches the equilibrium.

It is important to emphasize that  the large imbalance   in  eq.  (\ref{eq:V_difference})  is expressed in terms of $a_c$ which, by definition, represents the difference between initial value of $a(t=0)$ and finite value of $a(t_{\rm dec})$ at decoherence time $t_{\rm dec}$  when the axion field 
loses  its coherence. It does not depend on the rate  the axion field $\dot{a}(t)$  evolves, as discussed in Appendix  \ref{Appendix:CPodd_numerical}, where corresponding variations are parametrized by parameter $s_c$.
As one can see from  Fig.\ref{fig:numerical} the changes of the parameter $s_c$ do  not produce any visible modifications  in the final expression for disparity between nuggets and anti-nuggets. 

This is quite typical behaviour for all phenomena  related to the Berry's phase when the effects normally depend   on initial and final conditions rather than on specific dynamical properties of the system. The key element in all our discussions in this subsection is that we are interested in dimensionless ratio 
  (\ref{eq:V_difference}) between nuggets and anti-nuggets at the end of the evolution, when the sensitivity to all these dimensional parameters diminishes.
  The adiabatic approximation for the coherent axion field is also an important element in demonstration of   this ``insensitivity" of the final formula  (\ref{eq:V_difference}) to the parameters of the system.

\subsection{Survival of the  imbalance  between nuggets and anti-nuggets.   }\label{survival}
In the previous subsection we argued that  disparity (\ref{eq:V_difference}) which is generated during long lasting evolution of the coherent axion field is not sensitive to the initial conditions, nor to the axion mass $m_a$.
In this subsection we want to argue that the imbalance  (\ref{eq:V_difference}) survives 
the subsequent evolution of the system at $t>t_{\rm dec}$ after   the axion field loses  its coherence on the scale of the Universe as defined in eq. (\ref{a_c}), see also footnote \ref{coherence} with related comment. 

The basic argument is that the intensity of the axion field is drastically 
diminished  after $t_{\rm dec}$. However, the most important fact is not the amplitude of the remaining axion field, but rather its decoherence due to the 
emission of large number of randomly propagating axions with typical correlation length   $\lambda\sim  {\hbar}/{(m_a v_a)}$ rather than 
the Universe scale as in case (\ref{axion}). 
At this moment the coefficient $a\equiv 0$ in all our previous formulae. However, the   previously generated imbalance 
 (\ref{eq:V_difference})    
 does not disappear,  and it cannot be washed out because the coherent axion field (\ref{axion}) ceases to exist after $t_{\rm dec}$.
 In other words, the nuggets and anti-nuggets will continue to interact with environment by annihilating or accreting the quarks and anti-quarks from outside. 
 However, all these processes are not coherent on the scale of the Universe, and will equally influence both types of  species. 
 
We want to elaborate on another question which was previously mentioned in footnote \ref{a_large} and related to our technical assumption that 
$a_c<1 $   during the evolution  (\ref{eq:a}). In this case our treatment of $a(t)$ as a marginally small correction is justified.
However, our approach obviously breaks down when $a(t)$ becomes large.   There is nothing wrong in terms of physics of this evolution
with $a(t)>1$. It is just technical treatment of the problem which requires extra care.  

The increase $a(t)>1$ with time can be interpreted in terms of the original parameter $A\simeq\Delta\theta/\pi$ from  Eq. (\ref{eq:A2}) to become numerically large (close to one) as a result of accumulation of this phase during the long lasting influence of the background axion field. As we mentioned  in the text this $U(1)$ phase of the quark field is directly related to the baryon charge, see   \cite{Liang:2016tqc} for further technical details. When this phase becomes order of one the spectrum of states is completely reconstructed which effectively corresponds to the modification of the boundary conditions when the integer coefficient $N$ in formula (\ref{N}) is replaced by  $(N+1)$. After removing  the integer portion from the  $A$  to redefine $N$, the coefficient  $A\simeq\Delta\theta/\pi$ from  Eq. (\ref{eq:A2}) 
can be treated as a small parameter again. 
 
In all respects this procedure is   like conventional treatment of the angular field  $\phi (x)$ when $\phi(x)$ makes a full  cycle. 
The complete description of the system is accomplished by representing $\phi (x)$ in terms  of the integer number $N$ and fractional portion $0\leq \phi < 2\pi$ as a result of periodicity of the angular variable $\phi (x)$. If variable  $\phi(x)$ corresponds to a quantum field in quantum field theory (QFT) supporting the topological solitons, this analogy becomes very precise as parameter $N$ corresponds to a specific soliton  sector in this QFT model. 
The  fractional portion $\Delta\theta/\pi$ corresponds to the so-called fractionally charged  soliton, which is  well known construction in QFT, see   e.g. \cite{Liang:2016tqc} with references on the original literature in the given context. 
 
  For our specific problem on imbalance  between nuggets and anti-nuggets one should emphasize  that the  disparity  (\ref{eq:V_difference}) holds 
   irrespectively of the behaviour $a(t)$ during the evolution. If $a(t)$ becomes large at some moment of  the evolution, the corresponding portion of  $a(t)$ counts as a conventional        baryon charge $N$ in formula (\ref{N}) which obviously produces an additional imbalance between different species. 
   
   To summarize this section,  our main claim here is that the results obtained  above  (showing the disparity between nuggets and anti-nuggets)  are very robust in a sense that they are not very sensitive to the parameters of the theory, such as the axion  mass $m_a$ or the misalignment angle $\theta_0$.
   Furthermore, these results are not very sensitive to the detail behaviour of the system. We also argued that the generated imbalance cannot be washed out by consequent evolution of the system. Therefore, the computations of the present work strongly support  the assumption (\ref{ratio2}) which is 
   essentially equivalent to  eq. (\ref{Omega}), which is a generic  consequence of the framework.

\section{Conclusion and future development}\label{Conclusion}

 This work is a natural continuation of the previous  studies \cite{Liang:2016tqc}. 
 The crucial element which was postulated there  (without much computational support)  is represented  by eq. (\ref{ratio2}).
 In the present work, we investigate the evolution of domain wall bubbles in the presence of coherent $\cal CP$-odd axion field. We conclude that the coupling with the coherent axion  eventually leads to significant disparity (\ref{eq:V_difference})  in size between quark and antiquark nuggets. This provides an essential numerical and analytical support of eq. (\ref{ratio2}).  
 We summarize the main results and assumptions of present work as follows:

\begin{enumerate}
\item We assume the PQ transition happens before (or during)  the inflation, such that the vacuum is unique and the axion field $\theta$ is correlated on the scales of  the entire  Universe.  While the domain walls with $N_{\rm DW}>1$ cannot be formed in this case, 
 the so-called  $N_{\rm DW}=1$ domain walls, interpolating between one and the same physical vacuum, still can be formed,  as argued in \cite{Liang:2016tqc}.
This option has been overlooked somehow in previous studies because it had been previously  assumed that the all types of the domain walls   cannot  be  formed if the PQ transition happens before  the inflation. This element plays a key role in our analysis because the $\cal CP$-odd axion field is coherent on enormous scales of the entire Universe when the  $N_{\rm DW}=1$ domain wall bubbles can be also formed. 
 
 \item In the presence of a global coherent axion field $\theta$,  we argued  that a significant disparity (\ref{eq:V_difference})  between nuggets and 
 anti-nuggets will be generated.  In other words, we argued that the baryon charges separation effect inevitably occurs as a result of merely existence of the coherent axion field in early Universe. These studies essentially represent an explicit analytical and numerical support of the basic assumption (\ref{ratio2}) which is equivalent to the fundamental consequence of the framework (\ref{Omega}).

\item The accumulated disparity (\ref{eq:V_difference})   is  non-sensitive to the initial conditions, such as $\theta_{0}, T_0$, nor to the parameters of the system such as the axion mass $m_{a}$, as  argued in section \ref{sect:interpretation}. 

\item Furthermore, the imbalance  (\ref{eq:V_difference})  between nuggets and anti-nuggets
is not very sensitive to many dynamical parameters of the system. Rather, it is only sensitive to initial and final values of the  axion field when it  starts
to tilt at $t=0$ and  losses its coherence at moment $t=t_{\rm dec}$,  as argued in section \ref{survival}, see also relevant comment in footnote \ref{coherence}.  Such a behaviour  is, in fact,   a typical manifestation of the  accumulated    Berry's phase in condensed matter physics.     We also argued in section \ref{survival} that the subsequent evolution of the system cannot wash out the previously generated imbalance 
between nuggets and anti-nuggets.

\item
We avoid any  fine-tuning problems as a result of the features of the system listed   in items 3 and  4 above. It further supports our basic claim that the ratio $\Omega_{\rm dark} \approx \Omega_{\rm visible}$ is a very natural and universal outcome of this framework as  both types of matter (DM and visible)  are proportional to a single dimensional parameter of the system, $\Lambda_{\rm QCD}$. This claim  is  not sensitive to any specific details of the system.

 \item The ``baryogenesis" in this framework is replaced by ``charge separation" effect as reviewed in \cite{Liang:2016tqc}.  All, but one,   Sakharov's criteria \cite{Sakharov} are present in our framework (with exception of an explicit   baryon charge violation). Indeed, the  {\it $\cal{C}$ symmetry}    is broken    spontaneously       on the scale of an individual nugget   when the chemical potential $\mu$ (which is   odd value  under the $\cal{C}$ transformation)  is locally generated,    as explained in section \ref{spontaneous}. 
 The   {\it $\cal{CP}$ symmetry}    is broken globally as a result of the coherent axion field (which is odd under $\cal{CP}$ transformation) which generates the imbalance  (\ref{eq:V_difference})  between nuggets and anti-nuggets as highlighted in section \ref{CP}, and explained in great details in the present work. The generated   disparity (\ref{eq:V_difference}) cannot be   washed out at the later times  as a result of the {\it non-equilibrium} dynamics when the (originally coherent) axion field produces a large number of random  DM axions and loses its coherence at time $t_{\rm dec}$ as explained in section \ref{survival}.

\end{enumerate}

We want to conclude with few additional thoughts on the future directions within the  framework advocated in present work. 

It is quite obvious that a much deeper understanding of the QCD phase diagram at $\theta\neq0$ is essential   for any future progress, see Fig. \ref{fig:phase_diagram}. Due to the known ``sign problem'', the conventional lattice simulations cannot be used at $\theta\neq0$. The relevant recent studies \cite{D'Elia:2013eua,Bonati:2013tt,Bonati:2015uga,Bonati:2015sqt,Borsanyi:2016ksw} use a number of ``lattice tricks" to evade  the ``sign problem".  Still, the problem with a better understanding of the phase diagram at $\theta\neq 0$ remains. 

Another problem which is  worth to be mentioned is related to a deeper understanding of the  formation of closed domain-wall bubbles. Presently, very few results are available on this topic. The most relevant for our studies is the observation made in \cite{Sikivie:2008} that a small number of closed bubbles  are indeed observed in numerical simulations. However, their detail properties (their fate, size distribution, etc) have not been studied yet. A number of related questions such as an estimation of correlation length $\xi(T)$, the generation of the structure inside the domain walls, baryon charge accretion on the bubble, etc, hopefully can be also studied in such numerical simulations. 

One more possible direction for future studies from the ``wish list"   is  a  development of the  QCD-based  technique  related to  the evolution of the nuggets, cooling rates, evaporation rates, annihilation rates, viscosity, transmission/reflection coefficients, etc., in an unfriendly environment with non-vanishing $T,\mu,\theta$. All these and many other effects are, in general, equally contribute to our parameters like $T_{\rm form}$ and $c(T)$ at the $\Lambda_{\rm QCD}$ scale in strongly coupled QCD. Precisely these numerical factors eventually determine the coefficients in the observed relation Eq. (\ref{Omega}).

One more possible direction for future studies  from the ``wish list"  is  improvement of our current understanding by including the CS  gap in 
computations of the nugget's evolution. Such inclusion can result in a much more precise  restriction on the phenomenological parameter $c(T)$ in Eq. (\ref{ratio2}) relating  baryon to nugget ratio (\ref{ratio_omega}). 

As we mentioned in section \ref{intro}
this model    is consistent with all known astrophysical, cosmological, satellite and ground based constraints. The same  (anti)nuggets are also the source for the solar neutrino emissions.  A very modest improvement 
in the solar neutrino detection   may also lead to a discovery of the nuggets, see recent paper \cite{Lawson:2015cla}. 
One  can also argue that the same (anti)nuggets   may explain the long standing problem of the extreme UV and soft x-ray emission from 
the solar Corona \cite{Zhitnitsky:2017rop}.

Last but not least, the discovery of the axion  would conclude a long and fascinating journey of searches for this unique and amazing particle conjectured almost 40 years ago, see recent reviews   
  \cite{vanBibber:2006rb, Asztalos:2006kz,Sikivie:2008,Raffelt:2006cw,Sikivie:2009fv,Rosenberg:2015kxa,Graham:2015ouw,Ringwald2016}, and recent proposal \cite{Cao:2017ocv} on the axion search experiment which is sensitive to the axion amplitude $\theta$ itself, in contrast with conventional 
  proposals which are sensitive to $\dot{\theta}$. 
 
 If the  PQ symmetry    is broken before or during inflation (which is assumed to be the case in our framework as stated in item 1 at the beginning of this section)  then a  sufficiently large    axion mass   $m_a\gtrsim 10^{-4} {\rm eV}$ is unlikely to saturate the dark matter density observed today.   Indeed,  
in this case  the corresponding contribution to  $\Omega_{\rm dark}$ resulting from  the misalignment mechanism  \cite{misalignment}      is given by (see e.g. review \cite{Graham:2015ouw}):
 \be
 \label{dm_axion}
 \Omega_{\rm axion}\simeq \left(\frac{6\cdot 10^{-6} {\rm eV}}{m_a}\right)^{\frac{7}{6}}.
 \ee 
  This formula essentially states that the axion of mass $m_a\simeq 2\cdot 10^{-5}$ eV saturates the dark matter density observed today, while the axion mass in the range of  $m_a\gtrsim 10^{-4} {\rm eV}$ contributes very little to the dark matter density.  Formula   (\ref{dm_axion}) accounts only for the axions directly produced by the misalignment mechanism  and neglects the axions produced as a result of  decay of the topological defects which becomes the dominant mechanism if PQ phase transition occurs after the inflation\footnote{\label{defects}There is a number of uncertainties and  remaining discrepancies in the corresponding estimates.   We shall not comment on these  subtleties by referring to the   original   papers \cite{Chang:1998tb,Hiramatsu:2012gg,Kawasaki:2014sqa,Fleury:2015aca}. According to these estimates  
  the axion contribution to  $\Omega_{\rm dark}$  as a result of decay of the  topological objects   can saturate the observed DM density today if the axion mass is  in the range $m_a\sim10^{-4} {\rm eV}$.}.

  In the present work we advocate the idea that even if $m_a\gtrsim 10^{-4} {\rm eV}$ is large and the  PQ symmetry    is broken before or during inflation, still there is another complementary  mechanism contributing to $\Omega_{\rm dark}$ due to the ``quark nuggets'' formation.  We argue  that precisely this novel additional  mechanism     could provide    the principle contribution to dark matter of the Universe as the relation $\Omega_{\rm dark}\sim\Omega_{\rm visible}$ in this framework is not sensitive to the axion mass $m_a$, nor to the misalignment angle $\theta_0$ as advocated in the present work.
    
   \section*{Acknowledgments}
 
This work was supported in part by the National Science and Engineering
Research Council of Canada.

\appendix

  \section{\label{coherent}Coherent axion field (\ref{axion}) as the Berry's phase}
    In this Appendix   we shall argue that the  tiny phase difference proportional to $\theta(t)$ for quarks and antiquarks  trapped inside the nuggets and anti-nuggets
 might be interpreted as Berry's phase for the Fermi fields.  
 These  tiny changes  
 lead to small differences  in every individual QCD process    as shown in section \ref{sec:evolution}. 
 However, these small variations  are eventually translated into a large accumulated effect expressed in terms of global properties of the nuggets and anti-nuggets (such as their typical sizes). This   large effect formally expressed as (\ref{ratio2})  is  the direct consequence of the  very long  lasting accumulation of these  tiny $\cal CP$-odd effects due to fundamental  coherent   axion $\theta(t)$ field. Eventually,  the generation of $|c(T)|\sim1$ leads to the model-independent  prediction  of this framework expressed as (\ref{Omega}).  
 
 The   results of this adiabatic  evolution  are not sensitive to the parameters of the system, but rather   sensitive to the initial and final configurations of the system. Such a behaviour is very typical  for many phenomena  related to the accumulation of the Berry's phase in condensed matter  physics when the effects are  sensitive to the global rather than local characteristics of the system.
 Therefore, it is quite natural to expect that the behaviour of our system described  in sections    \ref{sec:evolution} and  \ref{difference} can be also interpreted as a result of accumulation of the Berry's phase which can be identified with the axion background field (\ref{axion}). 
 
  Our starting point is the $\theta$ term in QCD Lagrangian $\cal{L}_{\theta}$
  where $\theta (t)$ in this work is identified with the axion field
  \be
  \label{theta}
  {\cal{L}}_{\theta}=-\theta (t)\frac{g^2}{32 \pi^2}\tilde{G}_{\mu\nu}^a  {G}_{\mu\nu}^a.
  \ee
 As it is well known, one can  rotate the $\theta$ term away by rotating the quark fields.
 Assuming that we have $N_f$ light quarks with equal small masses one can represent this $U(1)_A$ chiral rotation
 in path integral formulation as follows 
 \be
 \label{phase}
 \psi\rightarrow \exp \left(-i\gamma_5 \frac{\theta(t)}{N_f}\right)\psi .
 \ee
 There is a number of important consequences of this transformation. We want to mention
 here just  few of them. 
 
 First of all, the phase (\ref{phase}) can be interpreted as the Berry's phase which is coherently accumulated 
 by all quark fields in entire Universe as long as field $\theta(t)$ given by (\ref{axion}) is coherent. This phase is obviously numerically much smaller than conventional phases related to the QCD fluctuations which are normally of order $\Lambda_{\rm QCD}$.
 However, the most important feature of this phase is that it is coherent on the scale of the entire Universe and can  be accumulated 
 during long period of time $\sim m_a^{-1}(t)$, much longer than a typical QCD processes with typical time scales $\sim  \Lambda^{-1}_{\rm QCD}$.

The Berry's phase entering (\ref{phase}) can be interpreted  as a result of acting of an auxiliary  (fictitious) magnetic field in the Hamiltonian $H_{\rm Berry}=\vec{\sigma}\cdot \vec{B}_{\rm Berry}$ where the so-called Berry's curvature $\vec{B}_{\rm Berry}$  
assumes the form 
\be
\label{B}
\vec{B}_{\rm Berry}\sim (m \cos\theta, ~~m\sin\theta, ~~\dot{\theta} ).
\ee
The  parameter $m$ enters the effective Lagrangian (\ref{eq:L4}) and has a physical meaning of a   QCD scale   as explained in the text.  The additional term $\sim \dot{\theta}$ in eq. (\ref{B}) is a  result of the $U(1)_A$ chiral rotation (\ref{phase}) in path integral when parameter of the rotation $\theta(t)$ depends on time.  The corresponding   Lagrangian generating this term has the  form $  {\cal{L}}_{5}=\dot{\theta}\bar{\psi}\gamma_0\gamma_5\psi$. The $\dot{\theta}$ in this expression  can be interpreted as 
axial chemical potential $\mu_5=\dot{\theta}$ which normally enters the Lagrangian as $ {\cal{L}}_{5}=\mu_5\bar{\psi}\gamma_0\gamma_5\psi$.

 In writing down an explicit expression for the Berry's curvature $\vec{B}_{\rm Berry}$ we used a specific  frame determined by   $\gamma_{\mu}$- representation as  given   in  \cite{Liang:2016tqc}. In this   representation 
 the auxiliary Berry's curvature $\vec{B}_{\rm Berry}$ can be thought as  a vector rotating along the equator in the $xy$ plane (neglecting small ${B}_{\rm Berry}^z\sim \dot{\theta}$ component) in this specific frame when time evolves. It is important to emphasize that this auxiliary field never returns to its original position after a complete cycle  because the  axion field reduces its amplitude  during the evolution. The corresponding energy of the coherent axion  field eventually goes to the production  of the propagating dark matter axions as a result of  conventional misalignment  mechanism. After the energy of the original field (\ref{axion}) is transferred to the propagating axions,  the large scale coherence (with the size  of the Universe) is lost, and the typical coherence length is determined by $\lambda_D$, which is much smaller scale. 
  
 The idea  that the axion field can be thought as an auxiliary (fictituous) magnetic field is not a new idea, and has been discussed in a number of papers in the past, see e.g. recent articles \cite{Graham:2013gfa,Stadnik:2013raa,Hill:2015kva,Barbieri:2016vwg} devoted to the axion  search experiments. Novel element advocating in the present work is that this field  (\ref{axion}) is coherent on enormous scales of the Universe before it decays to the propagating dark matter axions.  
 
 Another important consequence of accumulated phase (\ref{phase}) is that the ground state (the QCD vacuum) in the background 
 of the coherent axion field (\ref{axion}) explicitly violates $\cal{CP}$ invariance as can be explicitly seen by computing 
 the $\la {\rm det } \bar{\psi}_L^f \psi_R^f \ra$, see below.  The corresponding computations can be carried out in theoretically controllable way at sufficiently high temperature $T>T_c$ when the instanton approximation is justified. This region of temperatures  is precisely  when the domain wall network only starts to form and the axion field just starts to roll. 
 
 The corresponding technical computations are well known and presented in Appendix \ref{Appendix:det},
   \begin{eqnarray}
 \label{result}
\la {\rm det } ~\bar{\psi}_L^f \psi_R^f\ra  \sim  e^{i\theta(t)}  \cdot \Lambda_{QCD}^{3N_f}  \left(\frac{\Lambda_{QCD}}{T}\right)^{\frac{11}{3}(N - N_f)},~~~~~
   \end{eqnarray}
 where $f$ stands for flavour of a light quark.   Few comments about this important formula. First of all, the vacuum condensate (\ref{result}) does not vanish even in deconfined phase (well above the transition at $T\gg T_c$ shown on Fig. \ref{fig:phase_diagram}), in the region where  the chiral symmetry is restored and the chiral condensate itself  vanishes, i.e. $\la   \bar{\psi}  \psi \ra =0$.  This is because the vacuum condensate (\ref{result}) is formed  due to the explicit violation of the $U(1)_A$ symmetry rather than due to spontaneous $SU(3)_L\times SU(3)_R$ chiral symmetry. Furthermore, $\la {\rm det } ~\bar{\psi}_L^f \psi_R^f\ra$  does not vanish in the chiral limit $m_q\rightarrow 0$, see discussions in Appendix \ref{Appendix:det}.
   
   Another important comment is that formula  (\ref{result}) is derived in the dilute instanton gas approximation which is known to become a theoretically justifiable approximation   at sufficiently high temperature of order few times $\Lambda_{\rm QCD}$, see detail discussions on this matter in  Appendix \ref{Appendix:det}. Important comment we would like to make here is that the  decreasing  of the condensate  (\ref{result}) with  increasing the temperature 
   is much slower than in case of topological susceptibility  $\chi (T)\sim f_a^2 m_a^2 (T) $ which  determines  the axion mass dependence  on temperature $m_a(T)$, see recent numerical studies in refs.  \cite{Kitano:2015fla,Bonati:2015vqz,Borsanyi:2016ksw,Petreczky:2016vrs}. 
  
   Furthermore, the condensate (\ref{result}) also depends on the chemical potential (not shown explicitly in estimate  (\ref{result}) to simplify notations), 
   see  Appendix \ref{Appendix:det} with relevant comments. This feature has  an important implication  for our discussions  in   sections 
    \ref{sec:evolution} and  \ref{difference} as $\mu$-dependence of the vacuum condensate  (\ref{result}) implies   that the ground states inside and outside the nuggets would be   different.   
   
 The key  element for the present work is that the vacuum condensate (\ref{result}) explicitly depends on the axion phase $\theta(t)$ as a result of $U(1)_A$ transformation  (\ref{phase}). The presence of this phase unambiguously  implies that the ground state violates $\cal{CP}$ invariance
 even in the deconfined quark-gluon plasma well above the transition temperature $T_c$. This $\cal{CP}$  violation   occurred when $T>T_c$ which happened       long before the axion field    settles down at the origin $\theta=0$ when  the chiral phase transition occurs at $T\simeq T_c$. This remark has  some profound cosmological consequences   discussed in the main text in  sections     \ref{sec:evolution} and  \ref{difference}  
 as the phase (\ref{axion}) is correlated on enormous scales at this early stage of evolution.

\section{\label{Appendix:det} Computations  of the $\la {\rm det } \bar{\psi}_L^f \psi_R^f \ra$  }
The main goal of this Appendix is to derive formula (\ref{result}) for the vacuum expectation value $\la {\rm det } \bar{\psi}_L^f \psi_R^f \ra$ in deconfined phase where the instanton base computations are under complete theoretical control for sufficiently large $T$, 
which is precisely the region where the axion field starts to roll and the domain wall network starts to form. 
In context of the present work the generation of this condensate unambiguously  implies that the  ground state is $\cal{P}$ and  
$\cal{CP}$ odd  as a result of the   axion field $\theta$ given by (\ref{axion}) which is coherent on enormous scale at this period of the time evolution. One should emphasize that the $\theta$ dependence enters through the non-perturbative dynamics in deconfined regime.   As it is well-known the $\theta$ dependence cannot enter the dynamics on the level of   perturbation theory.

 We use the standard formula
for the instanton density at one-loop order \cite{tHooft,shuryak_rev,Rapp:1999qa}
\begin{eqnarray}
\label{instanton}
 n(\rho)&=& C_N(\beta(\rho))^{2N} \rho^{-5}
 \exp[-\beta(\rho)]  \nonumber\\
 &\times&  \exp[-(N_f \mu^2 + \frac13 (2N+N_f) \pi^2
 T^2)\rho^2], 
\end{eqnarray}
where
\begin{eqnarray}
\label{beta}
  C_N &=& \frac{0.466 e^{-1.679N} 1.34^{N_f}}{(N-1)!(N-2)! },~~~~\nonumber\\
\beta(\rho)&=&-b \log(\rho\Lambda_{QCD}), ~~~~  b=\frac{11}3 N-\frac23 N_f.  \nonumber
\end{eqnarray}
This formula  contains, of course, the standard instanton classical action $\exp(-8\pi^2/ g^2(\rho))
\sim  \exp[-\beta(\rho)]  $
which however is hidden as it is   expressed in terms of $\Lambda_{\rm QCD}$
rather than in terms of coupling constant $g^2(\rho)$.  This non-analytical dependence $\exp(-8\pi^2/ g^2)$ explicitly shows non-analytical and non-perturbative nature of the condensate $\la {\rm det } \bar{\psi}_L^f \psi_R^f \ra$ to be computed below based on expression (\ref{instanton}).  

We inserted the   chemical potential $\mu=\mu_B/N$ along with  temperature $T$ into this  expression  
to demonstrate that the instanton density becomes exponentially small for sufficiently large $T$ which explains the justification of 
the dilute instanton gas approximation in this regime. 

The computation of condensate $\la {\rm det } \bar{\psi}_L^f \psi_R^f \ra$  
is reduced to the following expression
\begin{eqnarray}
\label{det}
 \la {\rm det } \bar{\psi}_L^f \psi_R^f \ra=e^{i\theta(t)} \int d\rho n(\rho)d^4x\prod_i^{N_f}\frac{2\rho^3}{\pi^2
\left[x^2+\rho^2\right]^3} ~~~
\end{eqnarray}
 where we keep only zero modes in the chiral limit\footnote{It is known that the zero modes explicitly depend on the chemical potential, see e.g. review \cite{Rapp:1999qa}. We neglect these minor  corrections for our estimates in the present work.}, assuming that in the dilute gas approximation (which is  justified for sufficiently large $T$ as we mentioned above) all other mode contributions is suppressed by factor $m_q\rightarrow 0$. The integration over $d^4x$ corresponds to the integration over the instanton center at point $x$.
 
 Important point here is that the axion field $\theta$ explicitly enters the expression (\ref{det}), such that the condensate violates $\cal{P}$ and  
$\cal{CP}$ symmetries in the ground state. Another important comment is that the integral $\int d\rho$ is convergent, and for sufficiently large $T$ the expression (\ref{det}) represents the dominant contribution for this non-perturbative vacuum condensate. It is important to emphasize that the condensate $ \la {\rm det } \bar{\psi}_L^f \psi_R^f \ra$  does not vanish\footnote{In fact, this unique feature for this condensate  $ \la {\rm det } \bar{\psi}_L^f \psi_R^f \ra$ in the chiral limit  motivated a proposal to view the vacuum condense  $ \la {\rm det } \bar{\psi}_L^f \psi_R^f \ra$ as an order parameter to study the phase transition to the conformal window in the limit of large $N$ 
and finite $N_f/N\sim 1$, see \cite{Zhitnitsky:2013wfa} for references and details.}  in the chiral limit $m_q\rightarrow 0$, in contrast with partition function itself  $\cal{Z}$  which vanishes as ${\cal{Z}}\sim m_q^{N_f}$.  The  topological susceptibility $\chi\sim \partial^2 {\cal{Z}}/\partial \theta^2$, as well as the axion mass $m_a^2\sim \chi \sim m_q^{N_f}$  also vanish in the chiral limit\footnote{At low temperatures $T\simeq 0$ the corresponding features are quite different because the chiral condensate $\la\bar{\psi}_L^f \psi_R^f\ra\neq 0$, and also because in the $U(1)_A$ channel there are no massless degrees of freedom as the $\eta'$ is massive state.}. 
 
 After integration   over $d^4x$ one arrives to the following expression
 \begin{eqnarray}
 \label{det_1}
\la {\rm det }~ \bar{\psi}_L^f \psi_R^f\ra  =\frac{\pi^2\cdot e^{i\theta(t)} }
{(3N_f-1)(3N_f-2)} \int\!d\rho\, n(\rho) 
\rho^4 \biggl(\frac{2}{\pi^2 \rho^3} \biggr)^{N_f} , \nonumber
  \end{eqnarray}
where $n(\rho)$ is defined as before by eq.(\ref{instanton}). The combination  
 $ \int\!d\rho\, n(\rho) \rho^4$ is dimensionless while 
 the dimension of the operator $\la {\rm det }~
  \bar{\psi}_L^f \psi_R^f\ra\sim \la \rho\ra^{-3N_f}\sim$
  (MeV)$^{3N_f}$ as it should. After  integration over the instanton sizes $d\rho $ one arrives 
    \begin{eqnarray}
 \label{estimate}
\la {\rm det } ~\bar{\psi}_L^f \psi_R^f\ra  \sim  e^{i\theta(t)}  \Lambda_{\rm QCD}^{3N_f}  \left(\frac{\Lambda_{\rm QCD}}{T}\right)^{\frac{11}{3}(N - N_f)}
   \end{eqnarray}
  where for simple estimates we neglected all $(\log\rho)^n$ and all numerical factors  
  in   the integrand as they do not play any essential role in the  present work.     This formula is precisely the expression (\ref{result})  we used  in previous Appendix. 
    
\section{\label{technics}{Technical details}}
This appendix is devoted to derivation of  the pressure term given  by  Eq. (\ref{eq:P_simp}). It plays   an important role in our analysis  in section \ref{analytical}. The  basic idea is to use the  net baryon charge conservation given by  in Eq. (\ref{eq:evolution_B2}), and relate   the  radius of a domain wall bubble to its chemical potential, similar to the procedure we used in our previous studies in  \cite{Liang:2016tqc}. The new element now is the extra term $\sim a$ due to the background axion filed which has different signs for the nuggets and anti-nuggets. Th relevant formula reads, 
\be
x^2
&\simeq&\left(\frac{T_0}{T}\right)^2\frac{I^{(\pm)}_2(0)}{I^{(\pm)}_2(b)}
\left[1\mp\frac{AR_0T}{3\pi}x\cdot\frac{I^{(\pm)}_3(b)}{I^{(\pm)}_2(b)}\right]	\\
&\simeq&\left(\frac{T_0}{T}\right)^2\frac{I^{(\pm)}_2(0)}{I^{(\pm)}_2(b)}
\left[1\mp\frac{2}{3}ax
\left(\frac{3}{2}+\frac{2}{3}\sqrt{I^{(\pm)}_2(b)}\right)\right], \nonumber
\ee
where $x=R/R_{0}$, and $a=AR_{0}T/2\pi$ as defined in Eq. (\ref{eq:a}). In the second step, we use the approximated relation $I_3(b)/I_2(b)\simeq\frac{3}{2}+\frac{2}{3}\sqrt{I_2(b)}$, see Appendix \ref{Appendix:Fermi_integrals}. 

Our next step is to approximate and simplify  $I_2^{\pm}(b)$ by expanding it  with respect to small parameter $a$. We keep  the linear terms    only as $a$   is assumed to be numerically small parameter, i.e.
\be
\label{eq:B_simp2}
I_2^{(\pm)}[b(R)]
&\simeq&\left(\frac{T_{0}}{T}\right)^{2}\frac{I_{2}(0)}{x^2}
\left[1\mp\left(\frac{4}{9}\frac{T_{0}}{T}\sqrt{\frac{\pi^2}{12}}+x\right)a\right]  \nonumber\\
&\equiv& f^{(\pm)}(R).
\ee
We introduce a special notations for this combination   of $(R, a,  T)$ by introducing a function $f^{(\pm)}(R)$,  which enters  formula Eq. (\ref{eq:P_simp})
in the main text. This function is very useful because every higher order Fermi integrals $I_n(b)$ (and some simple polynomial functions) can be well approximated  in terms of  function of $I_2(b)$, see Appendix \ref{Appendix:Fermi_integrals}. This property allows us to rewrite all terms of chemical potential $b=|\mu|/T$ into simple function of radius $R$ using the relation (\ref{eq:B_simp2}). 

To conclude this Appendix we want to alert the readers that  the definition of $f^{(\pm)}(R)$ introduced above is slightly different from $f(R)$ introduced in our previous work  \cite{Liang:2016tqc}.    While the two functions   play a similar role in the analysis,   they are not identically the same even in the limit $a=0$. In the present work $f^{(\pm)}(R)$ is defined as $I_2[b(R)]$, while in the previous work it is defined as $f(R)\equiv I_2[b(R)]-\pi^2/6$. We opted to use a new definition (\ref{eq:B_simp2})
in the present  work because it produces much better accuracy with  the approximations and simplifications for the Fermi integrals adopted in the present work.
   In fact the accuracy of  present work is order $\pm5\%$ which should be compared with typical accuracy   $\sim20\%$ from the previous work \cite{Liang:2016tqc} when  similar approximations are made.

\section{\label{Appendix:CPodd_numerical} Numerical results  supporting (\ref{eq:V_difference}).}
Our goal here is to solve 
 the equation  ($\text{\ref{eq:R_simp2}}$) without using a large number of simplifications   and approximations of  Section $\text{\ref{sec:evolution}}$.
 Let us remind that the  goal of Section $\text{\ref{sec:evolution}}$ was  to make  a qualitative analysis leading to  (\ref{eq:V_difference}), rather than a precise quantitative description.
 In this Appendix  we use the exact form of Fermi integrals. In addition, we also keep the   contribution $\sim E_{B}\theta(b-b_{1})$ in (\ref{eq:P_simp}) 
 which was neglected in our qualitative analysis in Section $\text{\ref{sec:evolution}}$. 
 
 Furthermore, as we mentioned in Section  \ref{analytical}    we use the adiabatic approximation  in the computations of the nugget's dynamics (oscillations with slow damping). This adiabatic approximation is technically achieved by assuming that $m_a(T)$ and $T$ are  the  constants in the course of computations. 
 This assumption  can be only justified if the typical time scales of the relevant processes such as nugget's oscillations
are  much shorter than the time scale when the external parameters $[\theta(T), m_a(T),  T]$ vary. 
 To justify our approximation we   compute the following ratio $\omega_R/\omega_{\theta}$. 
 In this formula $\omega_R$ represents a typical frequency of the nuggets oscillation, while  $\omega_{\theta}$ 
 represents frequency of  oscillations of the axion field. 
 
 The computation of $\omega_{\theta}\simeq m_a$ is based on   interpolation formula (between low and high temperatures) for the topological susceptibility derived in   \cite{Wantz:2009it}:
 \be
 \label{m_axion}
 m_a^2f_a^2=1.46\cdot 10^{-3} \frac{ \Lambda^4(1+0.5\frac{T}{\Lambda})}{1+\left(3.53\frac{T}{\Lambda}\right)^{7.48}}, ~~~ \Lambda\simeq 0.4 ~{\rm GeV}.~~~~~~
 \ee 
 It is known that this expression deviates from the lattice results \cite{Kitano:2015fla,Bonati:2015vqz,Borsanyi:2016ksw,Petreczky:2016vrs}. Nevertheless,  it obviously reflects all the crucial elements in the behaviour of the topological susceptibility and the axion mass $m_a(T)$ as a function of the temperature $T$. It is certainly a sufficiently good approximation for our qualitative estimates of the ratio $\omega_R/\omega_{\theta}$.  
   
\begin{figure}
	\centering
	\captionsetup{justification=raggedright}
	\includegraphics[width=0.5\textwidth]{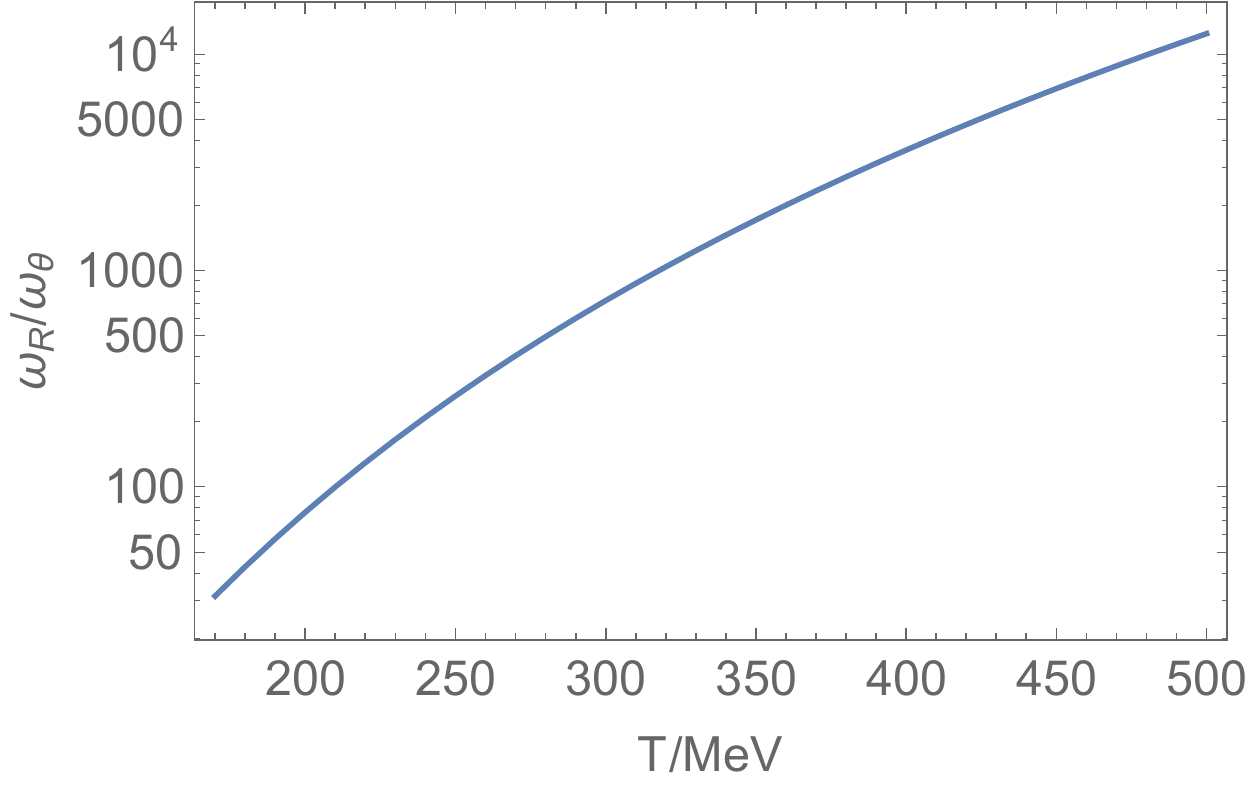}
	\caption{This plot shows that $\omega_R/\omega_{\theta}$ is always much larger than unity. This behaviour justifies our adiabatic approximation in numerical analysis when the axion mass $m_a (T)$ and the axion field $\theta(T)$ is  kept constant. This plot essentially shows that  the   nuggets make 
	  very large number of oscillations while the axion field $\theta(T)$ slowly varies.}
	\label{fig:ratio}
\end{figure}

 The computation of $\omega_R$ is based on numerical solution\footnote{In the corresponding numerical computations we do not assume the smallness of the oscillations. Therefore, we    do not use  expansion around the minimum of the potential leading to approximate linearized equation  (\ref{eq:6.R1}). } of the equation (\ref{eq:R_simp2}) for few consecutive oscillations of $R(t)$  for different values of the temperature $ 170~ {\rm MeV}\leq T\leq 500~ {\rm MeV}$.  The corresponding ratio $\omega_R (T)/\omega_{\theta}(T)$ as a function of the temperature is shown on Fig. \ref{fig:ratio}. The most important lesson from these computations is that this ratio is always much larger than unity, even in the vicinity of the chiral phase transition at $T\simeq 170~ {\rm MeV}$ when the chiral condensate forms and  the axion mass assumes its final maximum  value. This estimate unambiguously implies that our adiabatic approximation is justified as the nuggets make large number of oscillations while the axion field   varies only slightly as the relation $\omega_R\gg \omega_{\theta}$ states.

 As we shall  see below, the numerical studies of this Appendix  strongly support the qualitative analysis  presented in   Section $\text{\ref{sec:evolution}}$ and specifically the basic result (\ref{eq:V_difference}) demonstrating the disparity between nuggets and anti-nuggets due to the interaction with coherent axion field.

The parameter $a(t) $ was  introduced in eq. (\ref{eq:a}) to describe the accumulation of $\mathcal{CP}$-odd effects as the result of  
  evolution of the nuggets  in the background of axion field. As discussed in Sec. \ref{analytical}, $a(t)$ is a monotonically increasing function of time.
Essentially, this behaviour corresponds to the axion potential which  becomes more and more tilted with time as the axion mass $m_a(T)$ is increasing when the temperature slowly approaches the transition value $T_c$ from above.  When the axion potential becomes more tilted   the corresponding rate of the axion's emission   also increases. It obviously leads to the  decay of the axion amplitude $\theta$ as the energy from  the coherent axion field (\ref{axion}) is transferred to the propagating axions.   This effect of diminishing of the axion field  
is parametrized by $A$ in eq. (\ref{eq:A2}) which precisely describes the decreasing  of the axion amplitude during a single cycle. At some moment $\rm t_{\rm dec}$ in this evolution  the axion field  loses its coherence approaching its critical value $a_c$ as described in the text, see eq. (\ref{a_c}) and footnote \ref{coherence}. At this moment  the axion field may still produce  the axions, but it cannot  not lead  to any  coherent (on the scale of the Universe) changes  such as   (\ref{eq:V_difference}). 

In the numerical studies below, we concentrate  only on  computation of the disparity    (\ref{eq:V_difference})   while  the background approximation (\ref{eq:a}) remains valid. The generalization of this result on the case when the critical value $a_c$ becomes large
is straightforward and discussed in Section \ref{survival}. 

Therefore, we  model $a(t)$ to be  increasing  with time  from zero and (smoothly) stops at some cutoff value $a_c\leq1$ corresponding to the decoherence moment $t_{\rm dec}$, i.e. 
\begin{equation}
\label{eq:model_a}
a(s)=a_c\tanh\left(\frac{s}{s_c}\right),~~~~~~ s\equiv t/R_0
\end{equation}
where $a_c$ and $s_c$ serve as two free parameters. For convenience in numerical computation, we use the ``rescaled'' time $s\equiv t/R_0$ in this appendix.  The parameter $a_c$ is defined by  (\ref{a_c}) and corresponds to the moment when the axion field loses its coherence as explained above and  in footnote  \ref{coherence}.    The parameter $s_c$ describes  the rate the parameter $a(t)$ changes with time.  

\begin{figure*}
	\centering
	\captionsetup{justification=raggedright}
	\begin{subfigure}[t]{0.8\textwidth}
		\includegraphics[width=\textwidth]{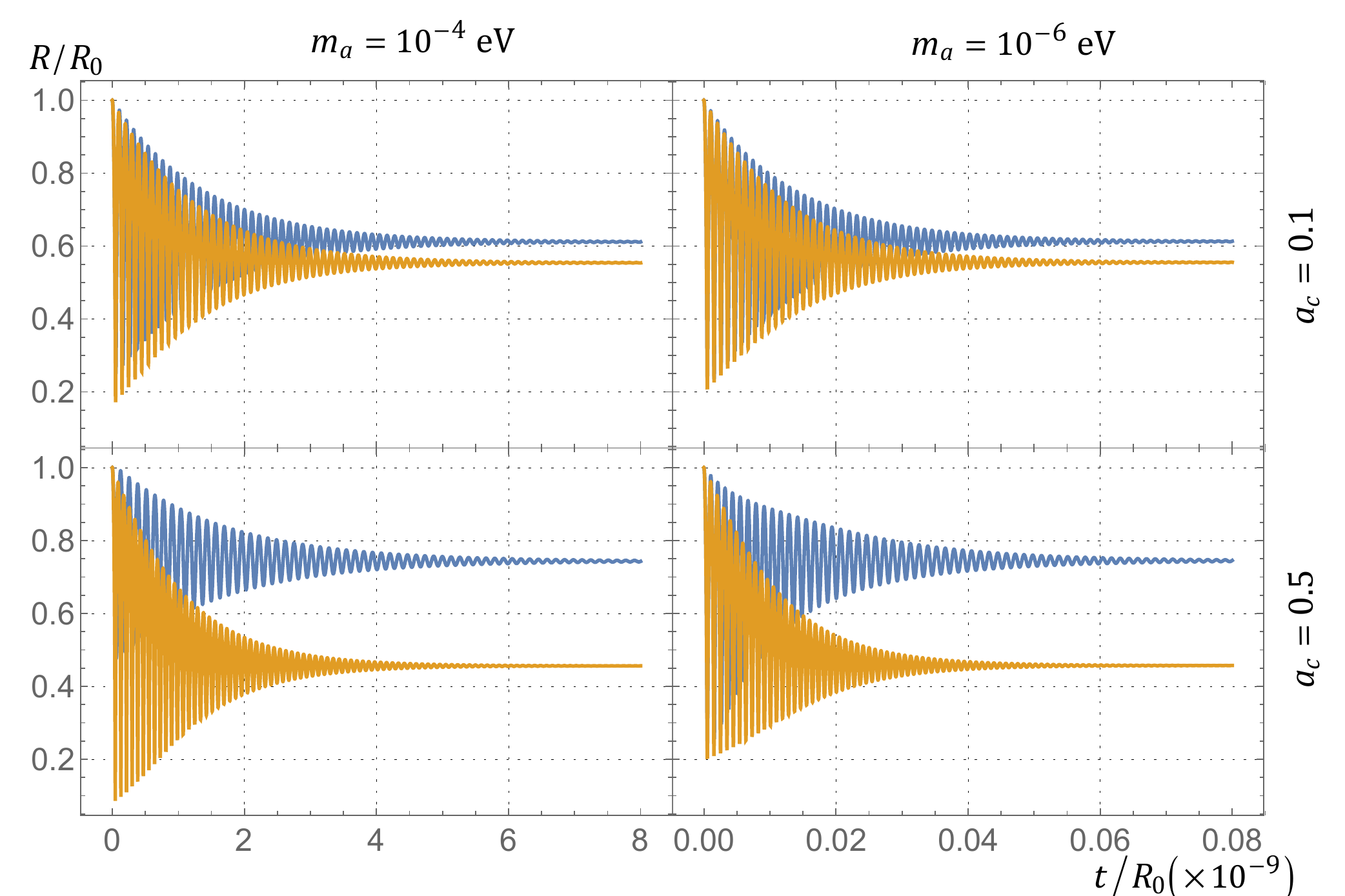}
		\caption{$s_{c}=10^{-2}$}
		\label{fig:numerical_a}
	\end{subfigure}
	\begin{subfigure}[t]{0.8\textwidth}
		\includegraphics[width=\textwidth]{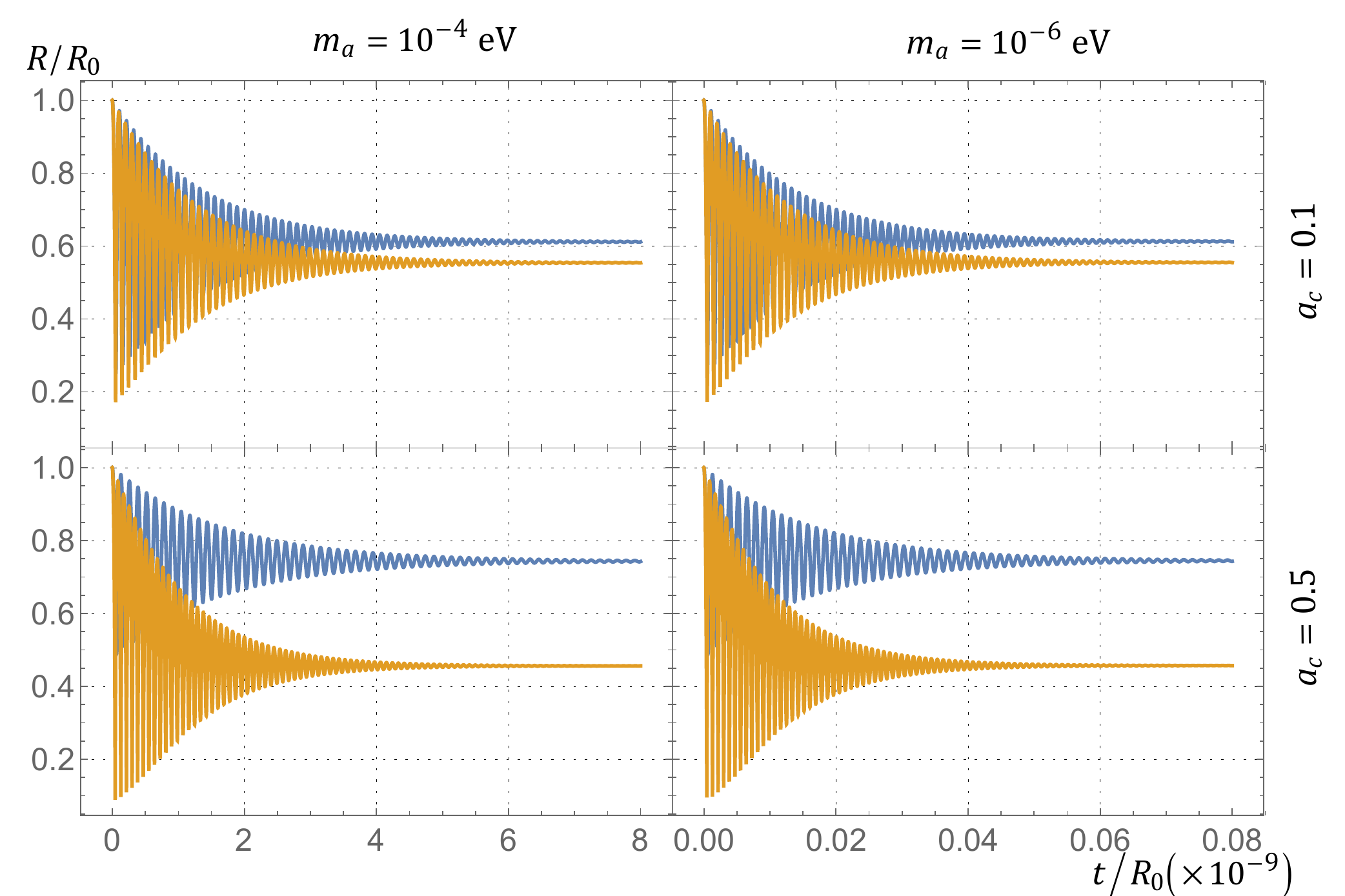}
		\caption{$s_{c}=10^{-5}$}
		\label{fig:numerical_b}
	\end{subfigure}
	\caption{Numerical solutions of (anti)nuggets evolving in the background of axion field before QCD transition. The blue and orange lines represent the evolutions of $R^{-}(s)$ and $R^{+}(s)$ respectively. We plot the upper four subfigures in (a) with $s_{c}=10^{-2}$ and the lower four subfigures in (b) with $s_{c}=10^{-5}$. The numerical values of parameters $m_a$ and $a_c$ that we use in calculating each subfigure can be seen in the upper edge and right edge of the graph.}
	\label{fig:numerical}
\end{figure*}

As manifested in numerical evaluation, the final   effects of the disparity  (\ref{eq:V_difference})   do not depend on what the exact model for $a(t)$ is. More specifically, the rate  of the variation of $a(t)$   does  not affect the final magnitude of the disparity between two species. Here we model it as the hyperbolic tangent form just for the convenience of numerical calculations.   In the model (\ref{eq:model_a}), we describe the critical  point using the cutoff $a_c$. As one  can  see from  Fig. \ref{fig:numerical}, the final disparity  effects do not depend on parameter $s_c$, but  only depend on the value of $a_c$ when the coherence of the axion field is lost. 
The Fig. \ref{fig:numerical} also shows that  the disparity effects  do not depend on the axion  mass $m_a$, 
in agreement   with  our arguments in Section \ref{sect:interpretation}. 

\begin{figure}
	\centering
	\captionsetup{justification=raggedright}
	\includegraphics[width=0.5\textwidth]{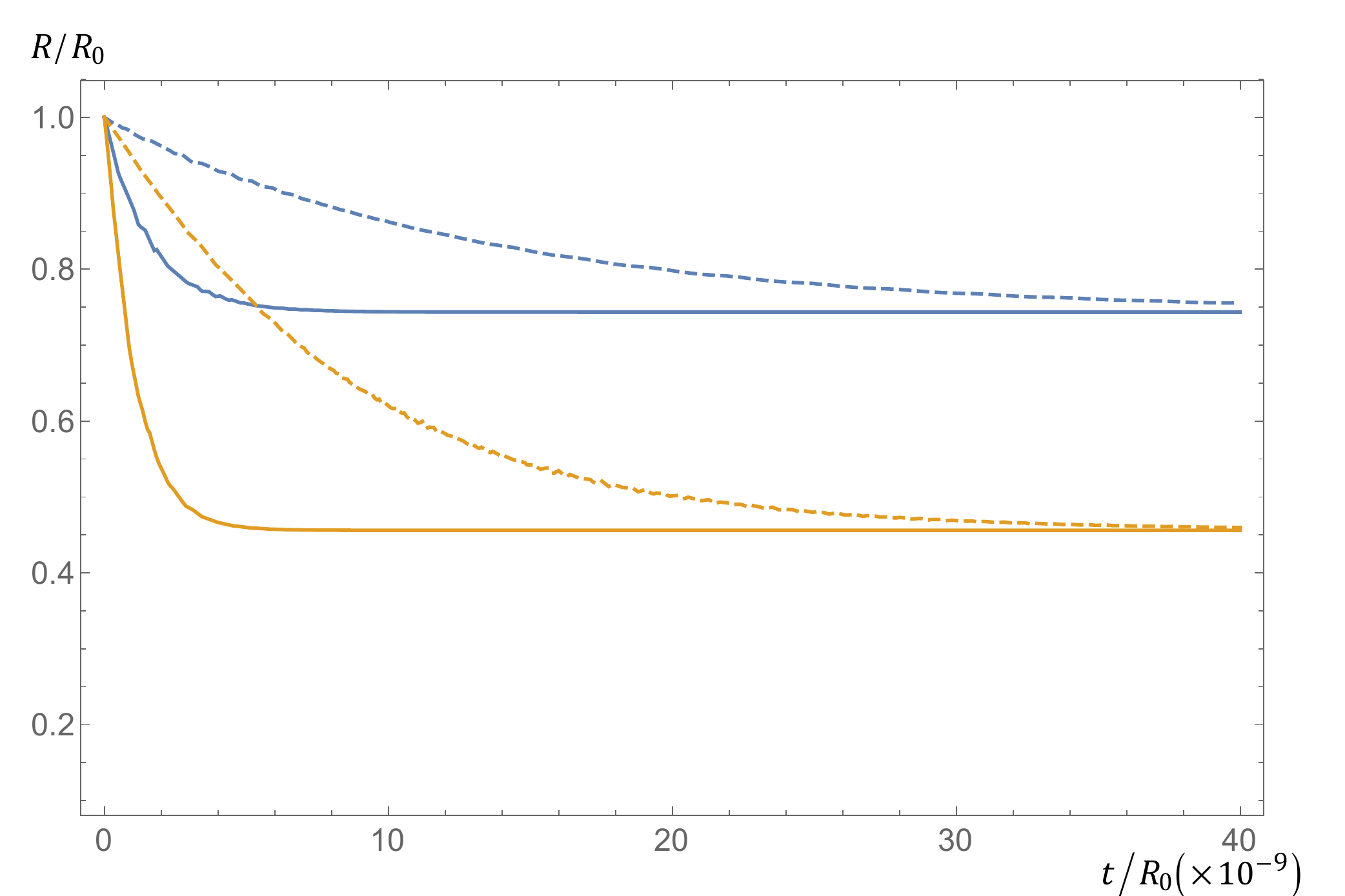}
	\caption{Dependence on viscosity $\eta$. Amplitudes of $R^-$ (blue) and $R^+$ (orange) are plotted. The solid lines correspond to $\eta=8.4m_\pi^3(\times10^9)$, and the dashed lines correspond to $\eta=1.0m_\pi^3(\times10^9)$. Here parameters $m_a=10^{-4}$ eV and $s_c=10^{-5}$ are chosen.}
	\label{fig:different_eta}	
\end{figure}

 The disparity effects  also  do not depend on the viscosity $\eta$.  To support this claim we did two different computations with different values of viscosity. The first choice is motivated by our previous studies  \cite{Liang:2016tqc} where we used $\eta\simeq m_{\pi}^3$ which has been computed in  different  models under different conditions in refs \cite{Arnold:2000dr,Chen:2007}. It is known that the viscosity is in fact is somewhat larger in the region of sufficiently  high temperature. Therefore, as the second choice we use the holographic arguments of ref. \cite{Kovtun2005} suggesting that $\eta$ could be one order of magnitude larger  than conventional perturbative QCD predicts (we use factor 8.4 in our numerical computations). 
One can explicitly see from Fig. \ref{fig:different_eta} that  the final destination for the disparity effects  do not  depend on the viscosity, in agreement with general arguments of Section   \ref{sect:interpretation}.

We summarize the numerical values of other parameters and constants needed in calculations in Table. \ref{table}. We choose initial temperature $T_{0}$ as $200$ MeV. The (anti)nuggets evolves in the background axion field from $200$ MeV to the QCD transition temperature $170$ MeV, and we can safely take the temperature as a constant, $T/T_{0}\simeq1$. 
\exclude{The QCD viscosity $\eta$ also needs some comments. The computations in \cite{Kovtun2005} suggest that $\eta/e\geq1/4\pi$ where $e$ is the entropy density. Here we set $\eta/e\simeq1/2\pi$, and using the relation of the outside pressure $P_{\rm out}\simeq\pi^2g^{\rm out}T^4/90$ and $e=\partial P/\partial T$, we have
\begin{equation}
\eta\simeq\frac{\pi g^{\rm out}}{45}T^3.
\end{equation}
Also, the QCD viscosity $\eta$ is a constant as $T$ is constant.
The observationally allowed window of axion mass is $10^{-6}{~\rm eV}\lesssim m_a\lesssim10^{-3}$ eV consistent with the recent constraints \cite{vanBibber:2006rb, Asztalos:2006kz,Sikivie:2008,Raffelt:2006cw,Sikivie:2009fv,Rosenberg:2015kxa,Graham:2015ouw,Ringwald2016}. To study the non-sensitivity to the initial conditions numerically, we should use different values of $m_a$. What we want to prove is that varying the value of $m_a$ will not affect the final magnitude of the $\mathcal{CP}$-odd effects. Here we choose $m_a\simeq10^{-4}$ eV and $10^{-6}$ eV respectively, and parameters $\sigma$, $R_0$ will be determined correspondingly, see Table. \ref{table}. 
}

\begin{table}[]
	\centering
	\caption{Table for some numerical parameters}
	\label{table}
	\begin{tabular}{lcc}
		\hline\hline
		Quantity & Value & \begin{tabular}[c]{@{}c@{}}QCD units\\ ($m_\pi=1$)\end{tabular} \\ \hline
		flavours $N_f$ & 2 & 2 \\
		colors $N_c$ & 3 & 3 \\
		degeneracy factor (in) $g^{\rm in} $ & 12 & 12 \\
		degeneracy factor (out) $g^{\rm out}$ & 37 & 37 \\
		
		bag constant $E_B$ & $(150~\rm{MeV})^4$ & $1.5$ \\
		``squeezer'' parameter $\mu_1$ & 330 MeV & 2.4 \\
		initial temperature $T_0$ & 200 MeV & 1.5 \\
		QCD viscosity $\eta$~ \cite{Arnold:2000dr,Chen:2007} & 0.002 GeV$^3$ & 1 \\
		QCD viscosity $\eta$~\cite{Kovtun2005} & 0.02 GeV$^3$ & 8.4 \\
		axion decay constant $f_a$ & $10^{10}$ GeV & $7.4\times10^{10}$ \\
		mass of axion $m_a$ & $10^{-4}$eV & $7\times10^{-13}$\\
		mass of axion $m_a$ & $10^{-6}$ eV & $7\times10^{-15}$\\
		domain wall tension $\sigma(m_a\sim 10^{-4}$eV) & $9\times10^{5}$ GeV$^3$& $3.7\times10^{8}$ \\
		domain wall tension $\sigma(m_a\sim 10^{-6}$eV) & $9\times10^7$ GeV$^3$& $3.7\times10^{10}$ \\
		initial radius $R_0\sim m_a^{-1}$ & $0.2$ cm & $1.4\times10^{12}$ \\
		initial radius $R_0\sim m_a^{-1}$ & $20$ cm & $1.4\times10^{14}$ \\
		\hline\hline
	\end{tabular}
\end{table}

\begin{figure}
	\centering
	\captionsetup{justification=raggedright}
	\includegraphics[width=0.5\textwidth]{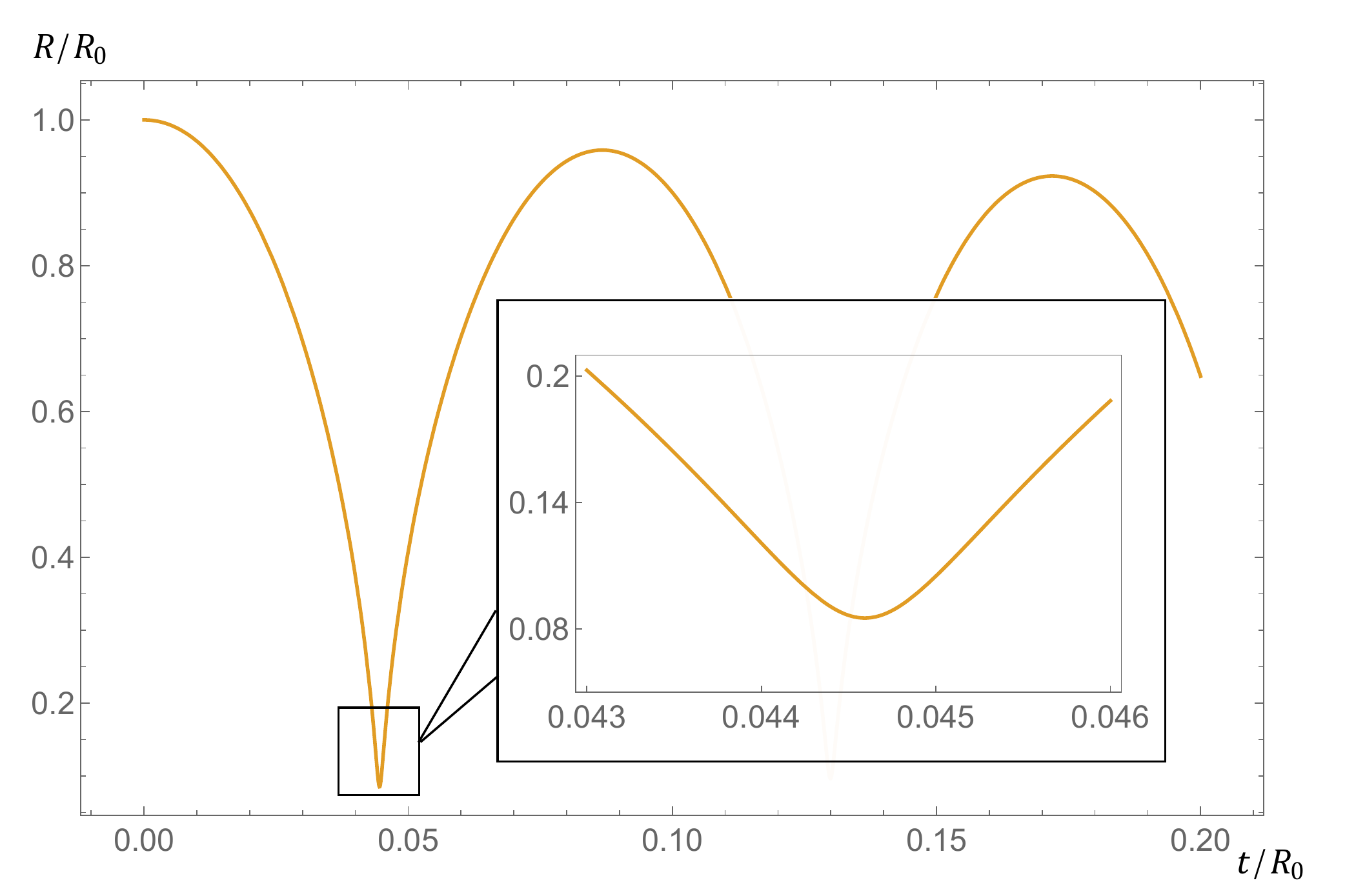}
	\caption{The first few oscillations of $R^{+}$ in the lower left subfigure of Fig. \ref{fig:numerical}. We choose this as an example to show that there is no cuspy problem.}
	\label{fig:cuspy}
\end{figure}
In Fig. \ref{fig:numerical_a}, we draw $2\times2$ subfigures
\footnote{To make the numerical computations solvable and the pattern of oscillations in Fig. \ref{fig:numerical} visible, here we adopt the rescaled QCD viscosity $\tilde{\eta}=10^9\eta$, following Ref. \cite{Liang:2016tqc}. This will not change any important results that we care about, like the formation radius of (anti)nuggets $R^{\pm}_{\rm form}$. The advantage is that this adoption will greatly ``shorten'' the evolution time and therefore make the numerical computations feasible. Correspondingly we add an extra factor $10^{-9}$ in the horizontal label in Fig. \ref{fig:numerical}. Otherwise if we directly use the value of $\eta$, Fig. \ref{fig:numerical} should be 9 orders of magnitude longer than shown.},
for $m_a\simeq(10^{-4}, ~10^{-6})$ eV, $a_c=(0.1, 0.5)$ and we set the parameter $s_c=10^{-2}$. We set the viscosity as $\eta=8.4~ m_{\pi}^3$. The Fig. \ref{fig:numerical_b} is the same as the Fig. \ref{fig:numerical_a} but with a different parameter $s_c=10^{-5}$. The blue and orange lines represent the evolution of $R^{-}(s)$ and $R^{+}(s)$ respectively. The difference between these two kinds of lines is the accumulated disparity effects. Comparing Fig. \ref{fig:numerical_a} with Fig. \ref{fig:numerical_b}, we see that changing $s_c$ will not affect the disparity effects. This verifies that the difference between two kinds of nuggets is non-sensitive to how $a$ increases, fast or slow. In Fig. \ref{fig:numerical_a} or Fig. \ref{fig:numerical_b}, comparing the four subfigures horizontally with $a_{c}$ fixed and $m_a$ varied, we see that the disparity effects are independent of $m_a$. This supports the arguments in Section \ref{sect:interpretation} that the disparity effects are non-sensitive to the mass of axion. Then we compare the four subfigures vertically with $a_{c}$ varied and $m_a$ fixed. We see that the disparity effects are determined by $a_c$ rather than other parameters. For $a_c=0.1$, we see that $\Delta R_{\rm form}=\left|R^{+}_{\rm form}-R^{-}_{\rm form}\right|\simeq 0.06$ and $\left<R_{\rm form}\right>=\frac{1}{2}\left|R^{+}_{\rm form}+R^{-}_{\rm form}\right|\simeq 0.6$, consistent with the analytical relation (\ref{eq:5C Rform_ratio}) in Section $\text{\ref{sec:evolution}}$. For $a_c=0.5$, $\Delta R_{\rm form}\simeq 0.3$ and $\left<R_{\rm form}\right>\simeq 0.6$, again consistent with the relation (\ref{eq:5C Rform_ratio}).

We also notice that the oscillations shown  on Fig. \ref{fig:numerical} are very sharp. But these seemingly cuspy behaviour is  in fact quite smooth on the QCD scale. To see this, we zoom in the first few oscillations of $R^{+}$ in the lower left subfigure of Fig. \ref{fig:numerical_a}, and plot it in Fig. \ref{fig:cuspy}. We see that the duration of ``cusp" is $\delta t_{\rm cusp}\sim10^{-3}R_0$, which is much longer than the QCD scale $\delta t_{\rm cusp}\gg\Lambda_{\rm QCD}^{-1}$. 
 One should also add that  the oscillation frequency is not sensitive to the viscosity $\eta$ according to eq. (\ref{eq:6.R1w}).  Therefore,  our comment about ``non-cuspy" behaviour remains unaffected as the time scale of a single oscillation (and therefore $\delta t_{\rm cusp}$) is not sensitive to the viscosity  $\eta$.

\section{\label{Appendix:Fermi_integrals} Fermi integrals}
We now study some more details on the Fermi integral of the following form:
\begin{equation}
I_n(b)=\int_0^\infty\frac{dx\cdot x^{n-1}}{e^{x-b}+1}.
\end{equation}
Such integrals can be exactly solved in terms of the so-called ``polylog'' function of degree $n$:
\begin{equation}
{\rm Li}_n(z)=\sum_{k=1}^{\infty}\frac{1}{k^n}z^k.
\end{equation}
Specifically, we obtain the following exact solution:
\begin{equation}
I_n(b)=-\Gamma(n){\rm Li}_n(-e^b).
\end{equation}
where $\Gamma(n)$ is the gamma function.
Also, one may find the polylog function satisfies following property when doing derivation:
\begin{equation}
\frac{d}{db}{\rm Li}_n(-e^b)={\rm Li}_{n-1}(-e^b).
\end{equation}
This property implies a useful relation between different $I_n$'s:
\begin{equation}
\frac{d}{db}I_n(b)=(n-1)I_{n-1}(b).
\end{equation}
For $b\geq0$, we sometimes prefer to approximate $I_n(b)$  in terms of ``basic'' functions:
\begin{subequations}
\begin{equation}
I_1(b)=\ln(1+e^b)
\quad\qquad\qquad\qquad\qquad({\rm Exact})
\end{equation}
\begin{equation}
I_2(b)\simeq\frac{\pi^2}{6}+\frac{1}{2}b^2-\frac{\pi^2}{12}e^{-b}
\qquad~\quad\qquad(\pm2\%)
\end{equation}
\begin{equation}
I_3(b)\simeq\frac{\pi^2}{3}b+\frac{1}{3}b^3+\frac{3}{2}\zeta(3)\cdot e^{-b}
\quad\qquad(\pm2\%)
\end{equation}
\begin{equation}
I_4(b)\simeq\frac{7\pi^4}{60}+\frac{\pi^2}{2}b^2+\frac{1}{4}b^4-\frac{7\pi^4}{120}e^{-b}
\quad(\pm3\%)
\end{equation}
\end{subequations}
All approximations are highly accurate within $\pm3\%$ uncertainty. Other useful approximations can be writing higher order $I_n(b)$ in terms of $I_2(b)$. For examples,
\begin{equation}
\frac{I_3(b)}{I_2(b)}\simeq \frac{3}{2}+\frac{2}{3}\sqrt{I_2(b)}
\qquad \qquad(\underbrace{\pm5\%}_{b\leq5};~\underbrace{\pm10\%}_{b\leq10}).
\end{equation}
Note that $\mu_{\rm max}\simeq500$ MeV is the critical upper limit of  QCD cutoff and $T\gtrsim170$ MeV before QCD transition, thus $b\lesssim3$ is the full applicable domain in this present work of study. A similar approximation for $I_4(b)$ is
\begin{equation}
I_4(b)\simeq 2\pi I_2(b)+(I_2(b))^2
\qquad\qquad\qquad(\underbrace{\pm 3\%}_{b\leq10}).
\end{equation}
Also, before transition, $T\gtrsim170$ MeV. Thus, this approximation is within $5\%$ of error since $b\lesssim3$. Also, near the transition $T\lesssim 220$ MeV, we can also approximate
\begin{equation}
b^2\simeq2.92+2I_2(b)
\qquad\qquad\qquad\quad\qquad(\underbrace{\pm3\%}_{b\geq1.5}),
\end{equation}
which is a valid approximation for $\mu>\mu_1\simeq330$ MeV.
 
 %

\newpage
 

\begin{thebibliography}{}

\bibitem{Zhitnitsky:2002qa} 
  A.~R.~Zhitnitsky,
  JCAP {\bf 0310}, 010 (2003)
  [hep-ph/0202161].
  
\bibitem{Oaknin:2003uv} 
  D.~H.~Oaknin and A.~Zhitnitsky,
  Phys.\ Rev.\ D {\bf 71}, 023519 (2005)
  [hep-ph/0309086].


\bibitem{Witten:1984rs}
E.~Witten,
  Phys.~Rev.~D {\bf 30}, 272 (1984).

\bibitem{Liang:2016tqc} 
  X.~Liang and A.~Zhitnitsky,
  Phys.\ Rev.\ D {\bf 94}, 083502 (2016)
  [arXiv:1606.00435 [hep-ph]].



\bibitem{Aoki:2006we} 
  Y.~Aoki, G.~Endrodi, Z.~Fodor, S.~D.~Katz and K.~K.~Szabo,
  Nature {\bf 443}, 675 (2006)
  [hep-lat/0611014].
  
  
  \bibitem{axion}
R.~D.~Peccei and H.~R.~Quinn,
Phys.\ Rev.\ D {\bf 16}, 1791 (1977);\\
S.~Weinberg,
Phys.\ Rev.\ Lett.\  {\bf 40}, 223 (1978);\\
F.~Wilczek,
Phys.\ Rev.\ Lett.\  {\bf 40}, 279 (1978).
\bibitem{KSVZ} J.E. Kim, Phys. Rev. Lett. {\bf 43} (1979) 103;\\
M.A. Shifman, A.I. Vainshtein, and V.I. Zakharov, Nucl. Phys. 
{\bf B166} (1980) 493(KSVZ-axion).
\bibitem{DFSZ}
 M. Dine, W. Fischler, and M. Srednicki, Phys. Lett. {\bf B104}
(1981) 199;\\
A.R. Zhitnitsky, Yad.Fiz. {\bf 31} (1980) 497; 
Sov. J. Nucl. Phys.
{\bf 31} (1980) 260 (DFSZ-axion). 
 \bibitem{vanBibber:2006rb}
  K.~van Bibber and L.~J.~Rosenberg,
  Phys.\ Today {\bf 59N8}, 30 (2006);


\bibitem{Asztalos:2006kz}
  S.~J.~Asztalos, L.~J.~Rosenberg, K.~van Bibber, P.~Sikivie, K.~Zioutas,
  Ann.\ Rev.\ Nucl.\ Part.\ Sci.\  {\bf 56}, 293-326 (2006).
  
\bibitem{Sikivie:2008}
  Pierre Sikivie,
  Lect.\ Notes Phys. {\bf 741}, 19 (2008)
  arXiv:0610440v2 [astro-ph].
 
\bibitem{Raffelt:2006cw} 
  G.~G.~Raffelt,
  Lect.\ Notes Phys.\  {\bf 741}, 51 (2008)
  [hep-ph/0611350].

\bibitem{Sikivie:2009fv} 
  P.~Sikivie,
  Int.\ J.\ Mod.\ Phys.\ A {\bf 25}, 554 (2010)
  [arXiv:0909.0949 [hep-ph]].

\bibitem{Rosenberg:2015kxa} 
  L.~J.~Rosenberg,
  Proc.\ Nat.\ Acad.\ Sci.\  (2015), 
  
    
\bibitem{Graham:2015ouw} 
  P.~W.~Graham, I.~G.~Irastorza, S.~K.~Lamoreaux, A.~Lindner and K.~A.~van Bibber,
  Ann.\ Rev.\ Nucl.\ Part.\ Sci.\  {\bf 65}, 485 (2015)
  [arXiv:1602.00039 [hep-ex]].
  
  \bibitem{Ringwald2016}
  A. Ringwald,
  arXiv \ preprint (2016)
  [arXiv:1612.08933 [hep-ph]]
 
  
\bibitem{Budker:2013hfa} 
  D.~Budker, P.~W.~Graham, M.~Ledbetter, S.~Rajendran and A.~Sushkov,
  Phys.\ Rev.\ X {\bf 4}, no. 2, 021030 (2014)
  [arXiv:1306.6089 [hep-ph]].
  
\bibitem{Graham:2013gfa} 
  P.~W.~Graham and S.~Rajendran,
  Phys.\ Rev.\ D {\bf 88}, 035023 (2013)
  [arXiv:1306.6088 [hep-ph]].
  
\bibitem{Rybka:2014cya} 
  G.~Rybka, A.~Wagner, A.~Brill, K.~Ramos, R.~Percival and K.~Patel,
  Phys.\ Rev.\ D {\bf 91}, no. 1, 011701 (2015)
  [arXiv:1403.3121 [physics.ins-det]].


  
\bibitem{Sikivie:2013laa} 
  P.~Sikivie, N.~Sullivan and D.~B.~Tanner,
  Phys.\ Rev.\ Lett.\  {\bf 112}, no. 13, 131301 (2014)
  [arXiv:1310.8545 [hep-ph]].
  
   \bibitem{Beck}
  C. Beck, 
   Phys.\ Rev.\ Lett.\  {\bf 111}, 231801 (2013),
   [arxiv:1309.3790[hep-ph]];
  
\bibitem{Stadnik:2013raa} 
  Y.~V.~Stadnik and V.~V.~Flambaum,
  Phys.\ Rev.\ D {\bf 89}, no. 4, 043522 (2014)
  [arXiv:1312.6667 [hep-ph]].
  
   
\bibitem{Sikivie:2014lha} 
  P.~Sikivie,
  Phys.\ Rev.\ Lett.\  {\bf 113}, no. 20, 201301 (2014)
  [arXiv:1409.2806 [hep-ph]].
  
\bibitem{McAllister:2015zcz} 
  B.~T.~McAllister, S.~R.~Parker and M.~E.~Tobar,
  Phys.\ Rev.\ Lett.\  {\bf 116}, no. 16, 161804 (2016)
  Erratum: [Phys.\ Rev.\ Lett.\  {\bf 117}, no. 15, 159901 (2016)]
  [arXiv:1512.05547 [hep-ph]].
  
  
  
\bibitem{Hill:2015kva} 
  C.~T.~Hill,
  Phys.\ Rev.\ D {\bf 91}, no. 11, 111702 (2015)
  [arXiv:1504.01295 [hep-ph]].
  
\bibitem{Hill:2015vma} 
  C.~T.~Hill,
  Phys.\ Rev.\ D {\bf 93}, no. 2, 025007 (2016)
  [arXiv:1508.04083 [hep-ph]].

\bibitem{Kahn:2016aff} 
  Y.~Kahn, B.~R.~Safdi and J.~Thaler,
  Phys.\ Rev.\ Lett.\  {\bf 117}, no. 14, 141801 (2016)
  [arXiv:1602.01086 [hep-ph]].
  
\bibitem{Barbieri:2016vwg} 
  R.~Barbieri {\it et al.},
  arXiv:1606.02201 [hep-ph].
  
\bibitem{Arvanitaki:2014dfa} 
  A.~Arvanitaki and A.~A.~Geraci,
  Phys.\ Rev.\ Lett.\  {\bf 113}, no. 16, 161801 (2014)
  [arXiv:1403.1290 [hep-ph]].
   

  
 
\bibitem{Kitano:2015fla} 
  R.~Kitano and N.~Yamada,
  JHEP {\bf 1510}, 136 (2015)
  [arXiv:1506.00370 [hep-ph]].
 
\bibitem{Bonati:2015vqz} 
  C.~Bonati, M.~D'Elia, M.~Mariti, G.~Martinelli, M.~Mesiti, F.~Negro, F.~Sanfilippo and G.~Villadoro,
  JHEP {\bf 1603}, 155 (2016)
  [arXiv:1512.06746 [hep-lat]].
  
\bibitem{Borsanyi:2016ksw} 
  S.~Borsanyi {\it et al.},
  Nature {\bf 539}, no. 7627, 69 (2016)
  [arXiv:1606.07494 [hep-lat]].
  
\bibitem{Petreczky:2016vrs} 
  P.~Petreczky, H.~P.~Schadler and S.~Sharma,
  Phys.\ Lett.\ B {\bf 762}, 498 (2016)
  [arXiv:1606.03145 [hep-lat]].
  
 
\bibitem{Zhitnitsky:2006vt} 
  A.~Zhitnitsky,
  Phys.\ Rev.\ D {\bf 74}, 043515 (2006)
  [astro-ph/0603064].

\bibitem{Zhitnitsky:2016cir} 
  A.~Zhitnitsky,
  EPJ Web Conf.\  {\bf 137}, 09014 (2017)
  doi:10.1051/epjconf/201713709014
  [arXiv:1611.05042 [hep-ph]].

   \exclude{


\bibitem{Oaknin:2004mn}
D.~H. Oaknin and A.~R. Zhitnitsky,
 Phys.~Rev.~Lett. {\bf 94}, 101301 (2005), arXiv:hep-ph/0406146.

\bibitem{Zhitnitsky:2006tu}
A.~Zhitnitsky,
  Phys.~Rev.~D {\bf 76}, 103518 (2007), arXiv:astro-ph/0607361.

  
\bibitem{Lawson:2007kp}
K.~Lawson and A.~R. Zhitnitsky,
  JCAP {\bf 0801}, 022 (2008), arXiv:0704.3064 [astro-ph].

\bibitem{Forbes:2009wg} 
  M.~M.~Forbes, K.~Lawson and A.~R.~Zhitnitsky,
  Phys.\ Rev.\ D {\bf 82}, 083510 (2010)
  [arXiv:0910.4541 [astro-ph.GA]].

\bibitem{Forbes:2006ba} 
  M.~M.~Forbes and A.~R.~Zhitnitsky,
  JCAP {\bf 0801}, 023 (2008)
  [astro-ph/0611506].
  
\bibitem{Forbes:2008uf} 
  M.~M.~Forbes and A.~R.~Zhitnitsky,
  Phys.\ Rev.\ D {\bf 78}, 083505 (2008)
  [arXiv:0802.3830 [astro-ph]].
  

\bibitem{Lawson:2012zu} 
  K.~Lawson and A.~R.~Zhitnitsky,
  Phys.\ Lett.\ B {\bf 724}, 17 (2013)
  [arXiv:1210.2400 [astro-ph.CO]].
  
\bibitem{Lawson:2015xsq} 
  K.~Lawson and A.~R.~Zhitnitsky,
  Phys.\ Lett.\ B {\bf 757}, 376 (2016)
  [arXiv:1510.02092 [astro-ph.GA]].
  
 
\bibitem{Lawson:2010uz} 
  K.~Lawson,
  Phys.\ Rev.\ D {\bf 83}, 103520 (2011)
  [arXiv:1011.3288 [astro-ph.HE]].
  

\bibitem{Lawson:2012vk} 
  K.~Lawson,
  Phys.\ Rev.\ D {\bf 88}, no. 4, 043519 (2013)
  [arXiv:1208.0042 [astro-ph.HE]].
  

\bibitem{Lawson:2013bya} 
  K.~Lawson and A.~R.~Zhitnitsky,
eProceedings for  Cosmic Frontier Workshop: Snowmass 2013 Menlo Park, USA, March 6-8,
2013, 
  [arXiv:1305.6318 [astro-ph.CO]].
 
\bibitem{Abers:2007ji} 
  E.~S.~Abers, A.~K.~Bhatia, D.~A.~Dicus, W.~W.~Repko, D.~C.~Rosenbaum and V.~L.~Teplitz,
  Phys.\ Rev.\ D {\bf 79}, 023513 (2009)
  [arXiv:0712.4300 [astro-ph]].


\bibitem{Gorham:2012hy} 
  P.~W.~Gorham,
  Phys.\ Rev.\ D {\bf 86}, 123005 (2012)
  [arXiv:1208.3697 [astro-ph.CO]].
  

\bibitem{Gorham:2015rfa} 
  P.~W.~Gorham and B.~J.~Rotter,
  arXiv:1507.03545 [astro-ph.CO].
  
\bibitem{Lawson:2015cla} 
  K.~Lawson and A.~R.~Zhitnitsky,
  arXiv:1510.07646 [astro-ph.HE].
  
}
 
\bibitem{Sikivie}
P.~Sikivie, 
Phys. Rev. Lett. {\bf 48} (1982) {1156};\\
A.~Vilenkin and A.E. Everett,  
Phys. Rev. Lett. {\bf 48} (1982) {1867}.

 \bibitem{Vilenkin}
A. Vilenkin, E.P.S. Shellard, ``Cosmic strings and other topological defects", 
Cambridge University Press, 1994 
  
    
  
\bibitem{Chang:1998tb} 
  S.~Chang, C.~Hagmann and P.~Sikivie,
  Phys.\ Rev.\ D {\bf 59}, 023505 (1999)
  [hep-ph/9807374].


\bibitem{FZ} 
M.~M.~Forbes and A.~R.~Zhitnitsky,
JHEP {\bf 0110}, 013 (2001) [arXiv:hep-ph/0008315].

\bibitem{SG}
G.~Gabadadze and M.A. Shifman, 
Phys. Rev.{\bf D 62}(2000){114003} [arXiv:hep-ph/0007345].

\bibitem{Son:2000fh} 
  D.~T.~Son, M.~A.~Stephanov and A.~R.~Zhitnitsky,
  Phys.\ Rev.\ Lett.\  {\bf 86}, 3955 (2001)
  [hep-ph/0012041].

\bibitem{KZ} T.W.B. Kibble, J. Phys. A9, 1387 (1976);\\
W.  Zurek, Nature {\bf 317}, 505 (1985)
\bibitem{KZ-review}
 W. Zurek, Phys. Rep. {\bf 276}, 177 (1996)
 
  
\bibitem{Wantz:2009it} 
  O.~Wantz and E.~P.~S.~Shellard,
  Phys.\ Rev.\ D {\bf 82}, 123508 (2010)
  [arXiv:0910.1066 [astro-ph.CO]].\\
  O.~Wantz and E.~P.~S.~Shellard,
  Nucl.\ Phys.\ B {\bf 829}, 110 (2010)
  [arXiv:0908.0324 [hep-ph]].
  
\bibitem{Hiramatsu:2012gg} 
  T.~Hiramatsu, M.~Kawasaki, K.~Saikawa and T.~Sekiguchi,
  Phys.\ Rev.\ D {\bf 85}, 105020 (2012)
  Erratum: [Phys.\ Rev.\ D {\bf 86}, 089902 (2012)]
  [arXiv:1202.5851 [hep-ph]].
  
\bibitem{Kawasaki:2014sqa} 
  M.~Kawasaki, K.~Saikawa and T.~Sekiguchi,
  Phys.\ Rev.\ D {\bf 91}, no. 6, 065014 (2015)
  [arXiv:1412.0789 [hep-ph]].
  
\bibitem{Fleury:2015aca} 
  L.~Fleury and G.~D.~Moore,
  JCAP {\bf 1601}, 004 (2016)
  [arXiv:1509.00026 [hep-ph]].
   
    
     \bibitem{misalignment}  
 J. Preskill, M. B. Wise, and F. Wilczek, Phys.Lett. {\bf B120}, 127 (1983);\\
  L. Abbott and P. Sikivie, Phys.Lett. B {\bf 120}, 133 (1983);\\
 M. Dine and W. Fischler, Phys.Lett. B {\bf 120}, 137 (1983).
 
\bibitem{Alford:2007xm} 
  M.~G.~Alford, A.~Schmitt, K.~Rajagopal and T.~Sch\"afer,
  Rev.\ Mod.\ Phys.\  {\bf 80}, 1455 (2008)
  [arXiv:0709.4635 [hep-ph]]
  
\bibitem{Rajagopal:2000wf} 
  K.~Rajagopal and F.~Wilczek,
  In *Shifman, M. (ed.): At the frontier of particle physics, vol. 3* 2061-2151
  [hep-ph/0011333].


       

\bibitem{Zhitnitsky:2013wfa} 
  A.~R.~Zhitnitsky,
  Nucl.\ Phys.\ A {\bf 921}, 1 (2014)
  [arXiv:1308.0020 [hep-ph]].
  
  
  
  
   
 
  \bibitem{Sakharov}
  A. D. Sakharov,   JETP Lett. {\bf 5}, 24 (1967).
  

\bibitem{D'Elia:2013eua} 
  M.~D'Elia and F.~Negro,
  Phys.\ Rev.\ D {\bf 88}, no. 3, 034503 (2013)
  [arXiv:1306.2919 [hep-lat]].
  
\bibitem{Bonati:2013tt} 
  C.~Bonati, M.~D'Elia, H.~Panagopoulos and E.~Vicari,
  Phys.\ Rev.\ Lett.\  {\bf 110}, no. 25, 252003 (2013)
  [arXiv:1301.7640 [hep-lat]].
  
\bibitem{Bonati:2015uga} 
  C.~Bonati,
  JHEP {\bf 1503}, 006 (2015)
  [arXiv:1501.01172 [hep-lat]].
  
\bibitem{Bonati:2015sqt} 
  C.~Bonati, M.~D'Elia and A.~Scapellato,
  Phys.\ Rev.\ D {\bf 93}, no. 2, 025028 (2016)
  [arXiv:1512.01544 [hep-lat]].
  
   
\bibitem{Lawson:2015cla} 
  K.~Lawson and A.~R.~Zhitnitsky,
  Phys.\ Rev.\ D {\bf 95}, no. 6, 063521 (2017)
  [arXiv:1510.07646 [astro-ph.HE]].

\bibitem{Zhitnitsky:2017rop} 
  A.~Zhitnitsky,
  arXiv:1707.03400 [astro-ph.SR].
 
 
\bibitem{Cao:2017ocv} 
  C.~Cao and A.~Zhitnitsky,
  Phys.\ Rev.\ D {\bf 96}, no. 1, 015013 (2017)
  [arXiv:1702.00012 [hep-ph]].
  
  
   \bibitem{tHooft}
G.~'t Hooft,
Phys.\ Rev.\  D {\bf 14}, 3432 (1976).

  \bibitem{shuryak_rev} T.~Schafer and E.~V.~Shuryak,
  Rev.\ Mod.\ Phys.\  {\bf 70}, 323 (1998).
  
\bibitem{Rapp:1999qa} 
  R.~Rapp, T.~Schäfer, E.~V.~Shuryak and M.~Velkovsky,
  Annals Phys.\  {\bf 280}, 35 (2000)
  [hep-ph/9904353].


\bibitem{Arnold:2000dr} 
  P.~B.~Arnold, G.~D.~Moore and L.~G.~Yaffe,
  JHEP {\bf 0011}, 001 (2000)
  [hep-ph/0010177].
 
   
\bibitem{Chen:2007}
  Jiunn-Wei Chen and Eiji Nakano.
   {  Phys.\ Lett.\ B} {\bf 647},  371  (2007).


  \bibitem{Kovtun2005} 
  P. K. Kovtun, D. T. Son, and A. O. Starinets,
  Phys.~Rev.~Lett. {\bf 94}, 111601 (2005) [arXiv:hep-th/0405231v2]
 
     
 


 
\end{thebibliography}
%
%

\end{document}